\documentclass[twocolumn]{webofc}

\usepackage[varg]{txfonts}   
\usepackage{bm}
\newcommand{\beq}{\begin{equation}}
\newcommand{\eeq}{\end{equation}}
\newcommand{\beqs}{\begin{eqnarray}}
\newcommand{\eeqs}{\end{eqnarray}}
\newcommand{\pvec}{\bm{p}}

\newcommand{\xvec}{\mathbf{x}}

\newcommand{\dvec}{\bm{d}}
\newcommand{\Pvec}{\bm{P}}

\newcommand{\Ecm}{E_{\rm cm}}

\newcommand{\me}[3]{\langle #1\vert\ #2\ \vert #3\rangle}
\newcommand{\qmagcm}{q_{\rm cm}}
\begin{document}
\title{Recent highlights with baryons from lattice QCD}
%
%

\author{\firstname{Colin} \lastname{Morningstar}\inst{1}
   \fnsep\thanks{\email{cmorning@andrew.cmu.edu}}}

\institute{Department of Physics, Carnegie Mellon University,
           Pittsburgh, PA 15213, USA}

\abstract{
Highlights from recent computations in lattice QCD involving baryons are presented.
Calculations of the proton mass and spin decompositions are discussed, a
percent level determination of the nucleon axial coupling is described, and
determinations of the proton and neutron electromagnetic form factors and
light-cone parton distribution functions are outlined.  
Recent results applying the so-called L\"{u}scher method to
meson-baryon systems are presented.  Key points emphasized are that much better precision
with disconnected diagrams is being achieved, incorporating multi-hadron
operators is now feasible, and more and more studies are being done with
physical quark masses.
}
\maketitle
\section{Introduction}
\label{intro}
Highlights from recent computations in lattice QCD involving baryons are presented
in this talk.  First, some introductory information about how baryons can be
studied in lattice QCD is presented. Calculations of the proton mass and spin 
decompositions are then discussed, a
percent level determination of the nucleon axial coupling is described, and
determinations of the proton and neutron electromagnetic form factors and
light-cone parton distribution functions are outlined.  A current approach, involving the
so-called L\"uscher method, to confronting the
challenge of studying baryon scattering and resonance properties in lattice QCD is
then discussed.  Some recent results applying the L\"uscher method to
meson-baryon systems are presented.  A new calculation of the timelike pion form factor is
highlighted to motivate future baryon form factor computations, especially the form
factors for the $\Delta$ baryon which will be important to neutrino experiments such as DUNE.
A new development with baryon-baryon interactions with
the HAL QCD method is highlighted, and an $H$-dibaryon warm up calculation
is presented.  Key points emphasized are that much better precision
with disconnected diagrams is being achieved, incorporating multi-hadron
operators is now feasible, and more and more studies are being done with
physical quark masses.

\section{Studying baryons in lattice QCD}

Finite-volume stationary-state energies are obtained in lattice QCD from 
Monte Carlo estimates of an $N\times N$ Hermitian correlation matrix 
   \beq C_{ij}(t)
   = \langle 0\vert\, O_i(t\!+\!t_0)\, \overline{O}_j(t_0)\ \vert 0\rangle.
   \eeq
Crucial to the success in extracting the energies is the use of
judiciously designed operators $\overline{O}_j$ to create the states of
interest
  \beq
    O_j(t)=O_j[\overline{\psi}(t),\psi(t),U(t)].
  \eeq
Temporal correlator estimates are obtained from path integrals over the quark 
$\psi,\overline{\psi}$ and gluon $U$ fields
  \[
  C_{ij}(t)= \frac{ \int {\cal D}(\overline{\psi},\psi,U)\ \ 
   O_i(t+t_0)\ \overline{O}_j(t_0)\ \ \exp\left(-S[\overline{\psi},\psi,U]\right)}{  
  \int {\cal D}(\overline{\psi},\psi,U)
 \ \exp\left(-S[\overline{\psi},\psi,U]\right)}
\]
which involve the QCD action in imaginary time, formulated on a space-time lattice
\beq
  S_{\rm QCD}[\overline{\psi},\psi,U] = \overline{\psi}\ K[U]\ \psi + S_G[U],
\eeq
where $K[U]$ is the quark Dirac matrix and $S_G[U]$ is the gluon action.
The integrals over the Grassmann-valued quark fields are done exactly,
bringing the  correlators into forms such as 
  \[
  C_{ij}(t)= \frac{ \int {\cal D}U\! \det K[U]
   f(K^{-1}[U],\cdots, K^{-1}[U]) e^{-S_G[U]}}{  
  \int {\cal D}U\ \det K[U]
  e^{-S_G[U]}}
  \]
where $K^{-1}[U]$ are the quark propagators in the gluon field $U$.
The remaining integrals over the gluon fields must be done with the
Monte Carlo method, which uses a Markov chain to generate a sequence of 
gauge-field configurations $ U_1, U_2,\dots, U_N$.
Including $\det K$ in the Monte Carlo updating and evaluating the
$K^{-1}$ in the numerator are the most computationally demanding parts
of the calculation usually.  A Metropolis method is employed with a sophisticated
global updating proposal, known as Rational Hybrid Monte Carlo (RHMC).

Usually many different hadron operators are needed, so it is efficient to assemble
them using basic building blocks, which we choose to be
covariantly-displaced LapH-smeared\cite{laph} quark fields:
  \beq
 q^A_{a\alpha j}= D^{(j)}\widetilde{\psi}_{a\alpha}^{(A)},
 \qquad  \overline{q}^A_{a\alpha j}
 = \widetilde{\overline{\psi}}_{a\alpha}^{(A)}
  \gamma_4\, D^{(j)\dagger}
  \eeq
where the LapH-smeared quark field is
   \beq
    \widetilde{\psi}_{a\alpha}(x) =
      {\cal S}_{ab}(x,y)\ \psi_{b\alpha}(y),
     \qquad {\cal S} = 
     \Theta\left(\sigma_s^2+\widetilde{\Delta}\right),
   \eeq
and each displacement $D^{(j)}$ is a product of smeared links:
\beq
 D^{(j)}(x,x^\prime) =
 \widetilde{U}_{j_1}(x)\ \widetilde{U}_{j_2}(x\!+\!d_1)
 \dots \widetilde{U}_{j_p}(x\!+\!d_{p-1})
  \delta_{x^\prime,\ x+d_{p}}.
\eeq
Above, the $a,b$ are color indices, $\alpha$ is a Dirac spin index,
$\sigma_s^2$ is the smearing cutoff, and $\widetilde{\Delta}$ is the
three-dimensional gauge-covariant Laplacian.
The gauge-covariant
displacements utilize stout link\cite{stout} variables
$\widetilde{U}_k(x)$.
To a good approximation, the LapH smearing operator is
  $
     {\cal S} = V_s V_s^\dagger
  $
where the columns of matrix $V_s$ are the eigenvectors
  of $\widetilde{\Delta}$ on each time slice.

\begin{figure}
\centering
 \includegraphics[width=3.5in]{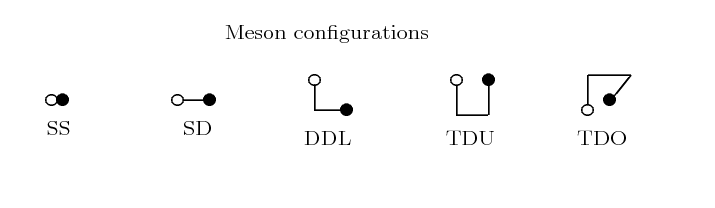}\\[-8mm]
 \includegraphics[width=3.2in]{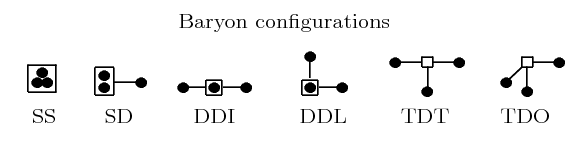}
 \caption{Spatial configurations of the meson and baryon operators
 in terms of displaced quark fields that we use.  SS stands for single site,
 SD for singly displaced, DDL for doubly displaced in an L configuration,
 TDO for triply displaced orthogonal, and so on.}
 \label{fig:hadronops}
\end{figure}

The quark displacements build up the orbital and radial structures
of the mesons and baryons.  The spatial configurations we use are
shown in Fig.~\ref{fig:hadronops}.
So-called elemental meson and baryon
operators are then given by
 \begin{align}
 \overline{\Phi}_{\alpha\beta}^{AB}(\pvec,t)&=
 \textstyle\sum_{\bm{x}} e^{i\pvec\cdot(\xvec+\frac{1}{2}(\bm{d}_\alpha+\bm{d}_\beta))}
   \delta_{ab}\ \overline{q}^B_{b\beta}(\bm{x},t)\ q^A_{a\alpha}(\bm{x},t),
 \\
  \overline{\Phi}_{\alpha\beta\gamma}^{ABC}(\pvec,t)&=
 \textstyle\sum_{\bm{x}} e^{i\pvec\cdot\xvec}\varepsilon_{abc}
\ \overline{q}^C_{c\gamma}(\bm{x},t)
\ \overline{q}^B_{b\beta}(\bm{x},t)
\ \overline{q}^A_{a\alpha}(\bm{x},t).
\end{align}
Group-theory projections onto the irreps of the lattice symmetry group
\beq
  \overline{M}_{l}(t)= c^{(l)\ast}_{
 \alpha\beta}\ \overline{\Phi}^{AB}_{\alpha\beta}(t),\qquad
  \overline{B}_{l}(t)= c^{(l)\ast}_{
 \alpha\beta\gamma}\ \overline{\Phi}^{ABC}_{\alpha\beta\gamma}(t),
\eeq
then produce meson $\overline{M}_{l}(t)$ and baryon $\overline{B}_{l}(t)$
operators which create states of definite momentum $\pvec$ 
in the irreps of the little group of $\pvec$.

The low-lying QCD mass spectrum has been successfully determined
(see, for example, Ref.~\cite{kronfeld}), with a
level of precision such that isospin breaking is now relevant.
The next challenge now is to evaluate scattering amplitudes and extract resonance
information.

\begin{figure}
\centering
\includegraphics[width=3.0in]{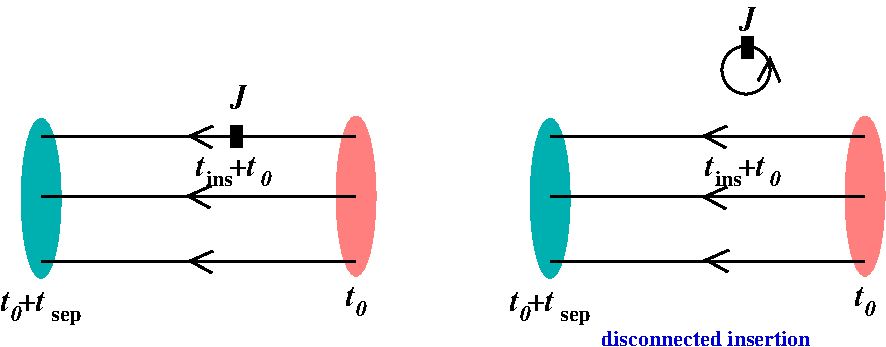}
\caption{Diagrammatic representations of contributions to 3-point functions.
(Left) The external current $J$ occurs on one of the three quark lines
connecting the baryon source (pink) to the baryon sink (blue).  (Right)
So-called disconnected contributions in which 
the insertion of the current $J$ does not connect to the three
quark lines between the baryon source and sink, but is involved in
a separate sea quark loop.
}
\label{fig:3ptfuncs}
\end{figure}

For matrix element calculations, the standard method requires evaluating
3-point functions, such as those shown in Fig.~\ref{fig:3ptfuncs}.  A major
issue is ensuring the removal of excited-state contamination by taking 
$t_{\rm sep},\ t_{\rm ins},$ and $t_{\rm sep}-t_{\rm ins}$ large.
In practice, this is difficult to achieve due to the signal-to-noise ratio which
decreases with time separation.  Computing so-called disconnected contributions,
as shown on the right in Fig.~\ref{fig:3ptfuncs}, is much more difficult
than the connected contributions, especially as the quark mass is taken to
its physical value.  Many previous studies simply neglected
such contributions.  However, a variety of new techniques are available now
and such disconnected contributions are being reliably estimated with
unprecedented precision.
Another issue with matrix element calculations is that the current operators require 
renormalization for comparison to $\overline{{\rm MS}}$. Such renormalizations
can be evaluated using nonperturbative and/or perturbative methods.

\begin{figure}
\centering
 \includegraphics[width=3.2in,clip]{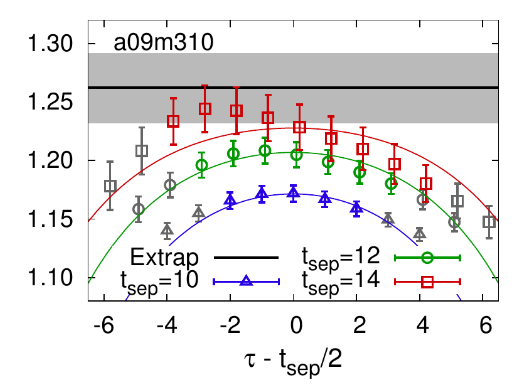}
 \caption{$g_A^{u-d}$ extractions for a variety of $t_{\rm sep}$, the time separation
 between the source and sink nucleon, and $ \tau=t_{\rm ins}$, the time where the
 current operator is inserted. Removal of excited-state contamination requires
 extrapolating to large $t_{\rm sep}$ and looking for insensitivity in $\tau$
 around  $t_{\rm sep}/2$.  Results are from Ref.~\cite{andre}.}
 \label{fig:3ptesc}
\end{figure}

Difficulties in removing excited-state
contamination in calculations related to $g_A$ are illustrated in 
Fig.~\ref{fig:3ptesc}.  The gray band shows the results of an extrapolation
of the $t_{\rm sep}=10,12,14$ results.  It is also necessary to ensure 
insensitivity of the extrapolated results to the insertion time, which must
be far from the source and sink.

\section{Proton mass decomposition}

\begin{figure}
\centering
\includegraphics[width=3.25in,clip]{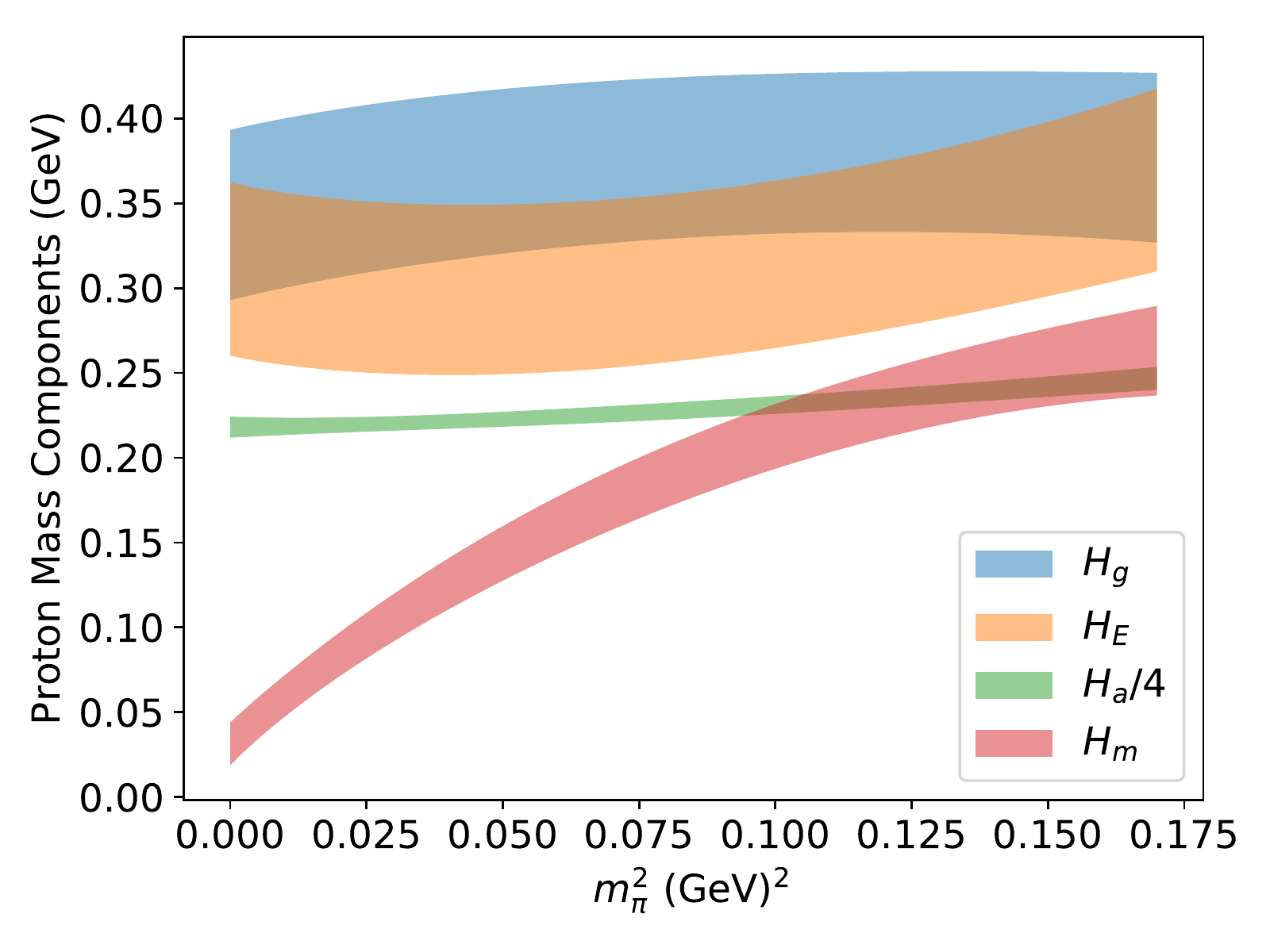}
\caption{Proton mass decomposition from Ref.~\cite{protonmass} against
the pion mass squared.  $H_g$ is the gluon field energy, $H_E$ is the
quark energy, $H_a$ is the anomaly term, and $H_m$ is the quark condensate.
}
\label{fig:protonmass}       
\end{figure}

A recent determination of the mass decomposition of the proton has been
presented in Ref.~\cite{protonmass}. The rest mass $M$ of the proton is given 
by\cite{JiProtonMass}
 \beq
 M = - \langle T_{44} \rangle =\langle H_m\rangle + \langle H_E\rangle (\mu)
+ \langle H_g\rangle (\mu) +  \textstyle\frac{1}{4} \langle H_a \rangle, 
\eeq
where $\langle T_{\mu\nu}\rangle$ is the expectation value of the
energy momentum tensor in a hadron, and
\begin{align}
 H_m\! &=\! \sum_{u,d,s\cdots}\int d^3x\, m\, 
\overline \psi  \psi ,\nonumber\\
 H_E\! &=\! \sum_{u,d,s...}\int d^3x~\overline \psi(\vec{D}\cdot \vec{\gamma})\psi,\nonumber\\
 H_g\! &=\! \int d^3 x~ {\frac{1}{2}}(B^2- E^2),\nonumber\\
H_a\! &=\! \sum_{u,d,s\cdots}\int d^3x\,\gamma_m m\, \overline \psi  \psi
  \!-\! \int d^3 x~ {\frac{\beta(g)}{g}}(E^2\!+\!B^2),\nonumber
\end{align}
where $H_m$ is the quark condensate, $H_E$ is the quark energy, $H_g$ is the gluon
field energy, and $H_a$ is the anomaly term.  Note that
 $\langle H_m\rangle$, $\langle H_a\rangle$, $\langle H_E+H_g\rangle$ are scale and scheme
 independent.  This study obtained the quark and gluon energies from the renormalized
quark and gluon momentum fractions
\[
 \langle H_g\rangle=\frac{3}{4}M\langle x\rangle_g,\qquad
 \langle H_E\rangle = \frac{3}{4}M\langle x\rangle_q-\frac{3}{4}\langle H_m\rangle,
\]
and the anomaly term from
 $\langle H_a\rangle = M -\langle H_m\rangle$.
The mass $M$ was determined from the two-point correlator, as usual,
and a previous determination of $\langle H_m\rangle$ by these authors was used.
The momentum fractions were evaluated using
\begin{align*}
\langle x \rangle_{q,g}\!&\equiv\!-\frac{\langle N |\frac{4}{3}\overline{T}^{q,g}_{44}
 |N \rangle}{M\langle N|N \rangle},\\
\overline{T}^{q}_{44}\!&=\!\int d^3x \overline{\psi}(x) \frac{1}{2}(\gamma_4
\overleftrightarrow{D}_4 -\frac{1}{4}{\displaystyle\sum_{i=0,1,2,3}} \gamma_i
\overleftrightarrow{D}_i) \psi (x), \nonumber\\ \ \ 
\overline{T}^{g}_{44}\!&=\!\int d^3x \frac{1}{2}(E(x)^2-B(x)^2),\nonumber
\end{align*}
taking renormalization into account:
\begin{align}
\langle x \rangle^R_{u,d,s}\!&=\!Z^{\overline{\textrm{MS}}}_{QQ}(\mu)\langle x 
\rangle_{u,d,s}+\delta Z^{\overline{\textrm{MS}}}_{QQ}(\mu)\!\!\!\sum_{q=u,d,s} 
\langle x \rangle_{q}+Z^{\overline{\textrm{MS}}}_{QG}(\mu)\langle x \rangle_g,
\nonumber \\
\langle x \rangle^R_g\!&=\!Z^{\overline{\textrm{MS}}}_{GQ}(\mu)\!\!\!\sum_{q=u,d,s} 
\langle x \rangle_{q}+Z^{\overline{\textrm{MS}}}_{GG}\langle x \rangle_g.
\nonumber
\end{align}
Results were obtained on four ensembles using an $N_{f}=2+1$ domain-wall fermion 
action with overlap valence propagators.  The difficult disconnected insertions
employed cluster-decomposition error reduction with all time slices looped over.
Extrapolations to remove systematic errors were done with a global fit including 
finite-volume and finite-spacing corrections, as well as known chiral 
behavior.  The impressive results are shown in Fig.~\ref{fig:protonmass}
and can be summarized as quark energy $32(4)(4)\%$, glue energy $36(5)(4)\%$,
quark condensate $9(2)(1)\%$, and trace anomaly $23(1)(1)\%$.

\section{Nucleon spin decomposition}
\begin{figure}
\includegraphics[width=1.5in]{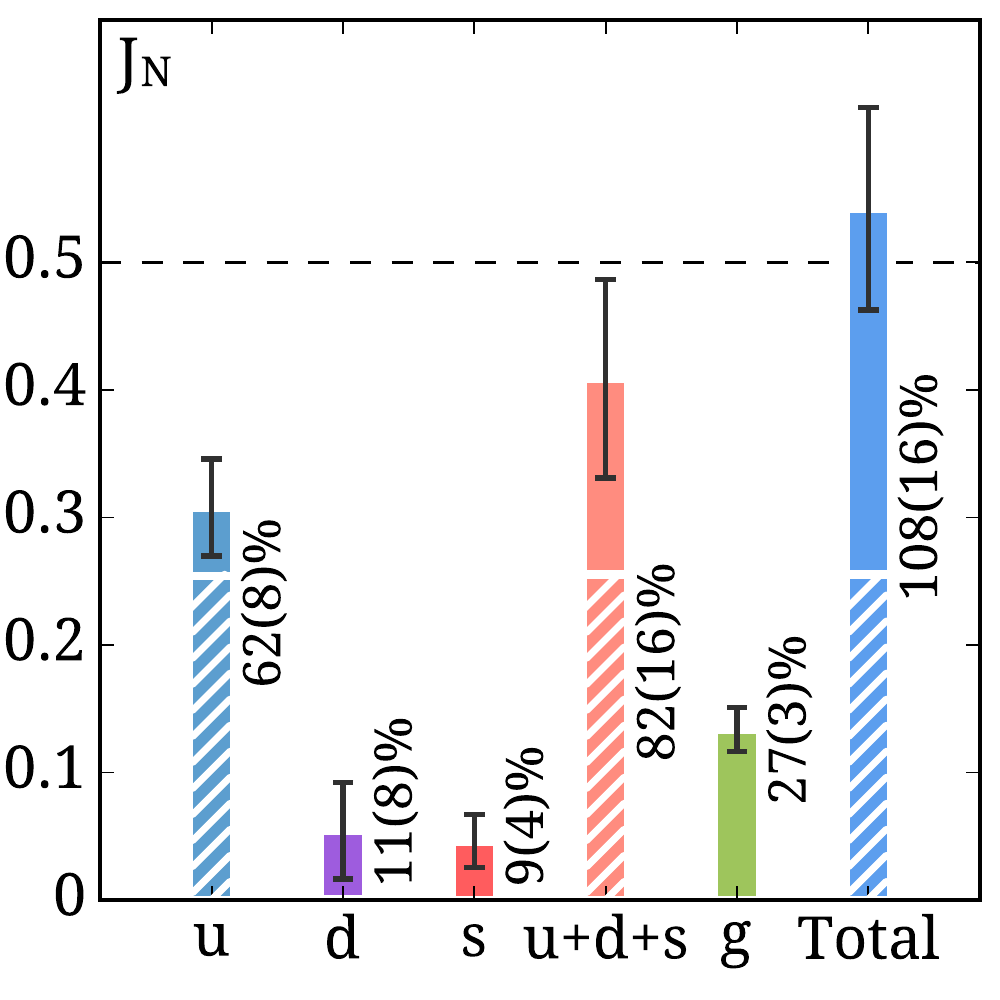}\hspace{3mm}
\includegraphics[width=1.5in]{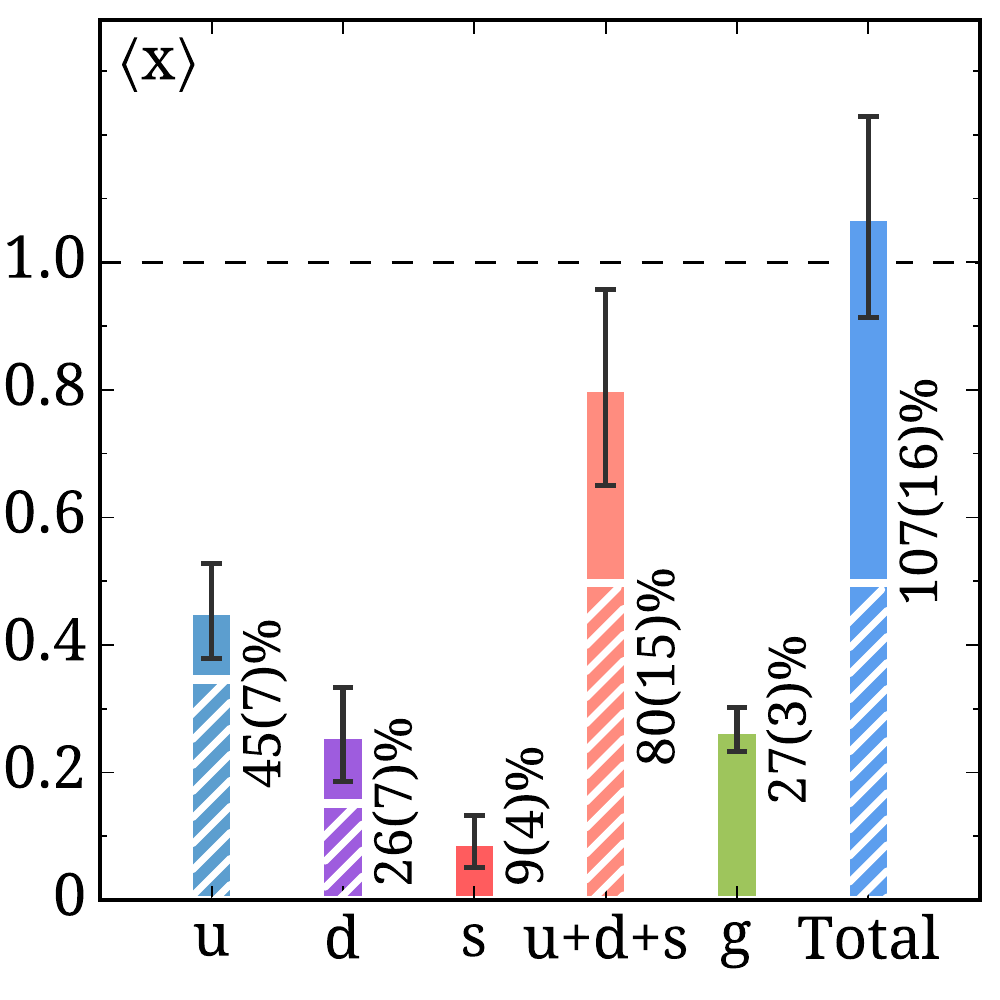}
\caption{Nucleon spin (left) and momentum fraction (right) decompositions
 from Ref.~\cite{protonspin} in terms of the contributions from each quark flavor
and from gluons. Striped segments refer to valence quark contributions,
while solid segments show sea quark and gluon contributions. }
\label{fig:spindecomp}
\end{figure}

A recent determination of the spin and momentum fraction
decomposition of the nucleon has been
presented in Ref.~\cite{protonspin}. 
From the Ji sum rule\cite{JiProtonSpin}, the nucleon spin is given by
\beq
 J_{N}{=} \sum_{q{=}u,d,s,c\cdots}\bigg(\textstyle\frac{1}{2}\Delta\Sigma_q + L_q \bigg)+
J_g{=} \sum_{q}J_q+
J_g,
\eeq
where $J_g$ is the gluon total angular momentum, $L_q$ is the quark orbital angular
momentum, and $\frac{1}{2}\Delta\Sigma_q$ is the contribution from the intrinsic
quark spin.  These quantities were obtained from the following nucleon matrix elements,
with $Q{=}p^\prime{-}p$ and $P{=}\frac{1}{2}(p^\prime{+}p)$:
\begin{align}
  \langle N(p,s^\prime)|\bar{q}\gamma_\mu \gamma_5 q|N(p,s)\rangle &{=} \bar{u}_N(p,s^\prime) \Bigl[g_A^q\gamma^\mu\gamma_5\Bigr]u_N(p,s),\nonumber\\
  \langle N(p^\prime,s^\prime)|\bar{q}\gamma^{\{\mu}\overleftrightarrow{D}^{\nu\}} q| N(p,s)\rangle &{=} \bar{u}_N(p^\prime,s^\prime)\Lambda^q_{\mu\nu}(Q^2)u_N(p,s),\nonumber\\
  \Lambda_{q(g)}^{\mu\nu}(Q^2) {=}  A^{q(g)}_{20}(Q^2) \gamma^{\{ \mu}P^{\nu\}} 
  &+B^{q(g)}_{20}(Q^2) \frac{\sigma^{\{ \mu\alpha}q_{\alpha}P^{\nu\}}}{2m}\nonumber\\
  &+C^{q(g)}_{20}(Q^2)\, \frac{1}{m}Q^{\{\mu}Q^{\nu\}}.\nonumber
\end{align}
The quark(gluon) total angular momentum and quark momentum fraction
 and spin were extracted using
\begin{eqnarray*}
 J_{q(g)}&{=}&\textstyle\frac{1}{2}[A_{20}^{q(g)}(0)+B_{20}^{q(g)}(0)],\\
 \langle x\rangle_{q} &=& A^{q}_{20}(0),\qquad
 \Delta\Sigma_q =g_A^q,
\end{eqnarray*}
and the gluon momentum fraction was obtained from ${\cal {O}}^g_{\mu\nu}{=}2 {\rm
  Tr}[G_{\mu\sigma}G_{\nu\sigma}]$ with
$\overline{\cal{O}}^g{\equiv} {\cal {O}}^g_{44}-\frac{1}{3}{\cal{O}}^g_{jj}$,
and
\[
  \langle N(p,s^\prime)|\overline{\mathcal{O}}^g|N(p,s)\rangle {=} \biggr 
  (-4E_N^2-\frac{2}{3}\vec{p}^2\biggl) \langle x\rangle_g.
\]
One ensemble at the physical point on a $48^3\times 96$ lattice using a twisted mass
clover-improved action with lattice spacing $a=0.0939(3)~{\rm fm}$, set from the nucleon mass,
was used.  The $u,d$ disconnected diagrams were estimated by exact deflation plus the
one-end-trick, while the $s$ disconnected diagrams were evaluated by a truncated solver 
method.  Renormalization factors were determined nonperturbatively.
Their final results are shown in Fig.~\ref{fig:spindecomp}.

\section{Nucleon axial coupling}

\begin{figure}
\centering
\includegraphics[width=3.25in]{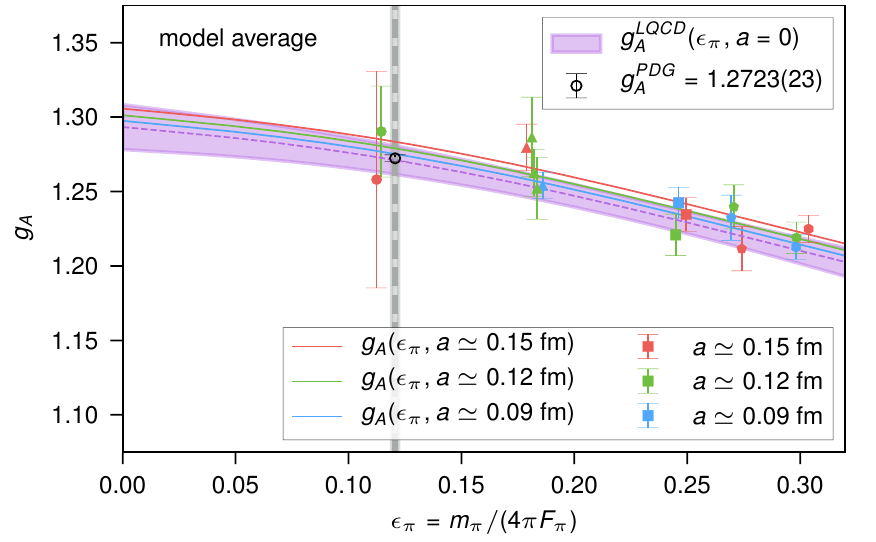}
\caption{Extrapolation of a $g_A$ determination in Ref.~\cite{andre} to the physical point.  
Solid red, green,
and blue curves are central values of $g_A$ as a function of $\epsilon_\pi=m_\pi/(4\pi F_\pi)$
at fixed lattice spacing and infinite volume, and the black circle is the experimental
value.  Magenta band is the central 68\% confidence band of the continuum and infinite
volume extrapolations.}
\label{fig:gAplot}
\end{figure}

\begin{figure}
\centering
\includegraphics[width=3.20in]{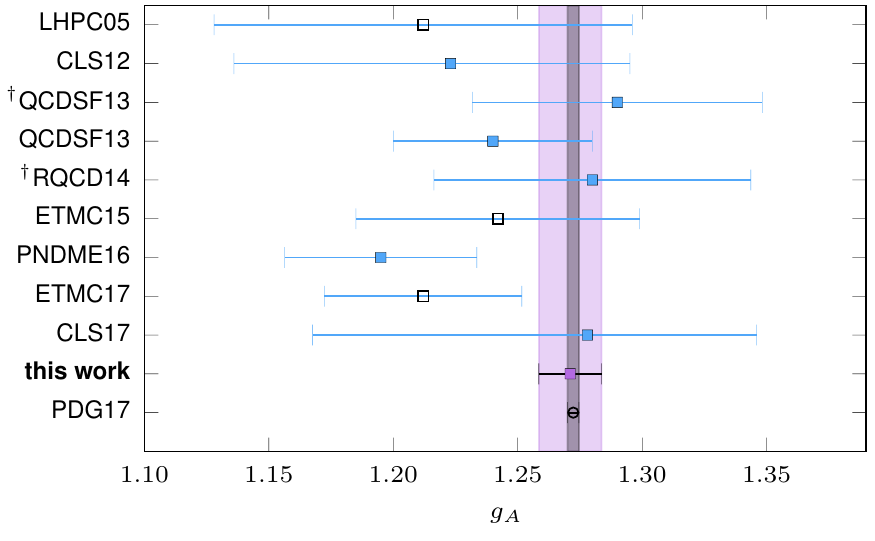}
\caption{Comparison of the estimate of $g_A$ from Ref.~\cite{andre} with other
recent determinations.  See Ref.~\cite{andre} for references to the other estimates. 
Results with closed symbols have included an extrapolation to the continuum limit,
while results with open symbols have only included extrapolation to the physical
pion mass.  To guide the eye, the vertical magenta band is the full uncertainty from 
Ref.~\cite{andre},
while the vertical gray band is the experimental uncertainty.}
\label{fig:gAcompare}
\end{figure}

A remarkable percent level determination of the nucleon axial coupling
$g_A$ in lattice QCD recently appeared\cite{andre}.
The use of a Feynman-Hellman method enabled a significant reduction in statistical
errors.  Their value was
 \beq
   g_A = 1.2711(103)^s(39)^\chi(15)^a(19)^V(04)^I(55)^M,
 \eeq
where the errors, in order, are from the statistical estimation, chiral extrapolation, 
lattice spacing extrapolation, volume extrapolations, isospin corrections, and
model selection.  Extrapolations were done using several models and the final
estimate is a model average.  Sixteen ensembles from the MILC collaboration
generated using a HISQ action with lattice spacings $a\sim 0.15$~fm,
$a\sim 0.12$~fm, and $a\sim 0.09$~fm and pion masses ranging from near physical to
400~MeV were used.  Domain-wall valence propagators were used, making this a mixed-action
computation.  Their main result is illustrated in Fig.~\ref{fig:gAplot},
and a comparison of their estimate with other recent determinations is shown
in Fig.~\ref{fig:gAcompare}.

\section{Proton/neutron electromagnetic form factors}
\begin{figure*}
\centering
\includegraphics[width=3.0in]{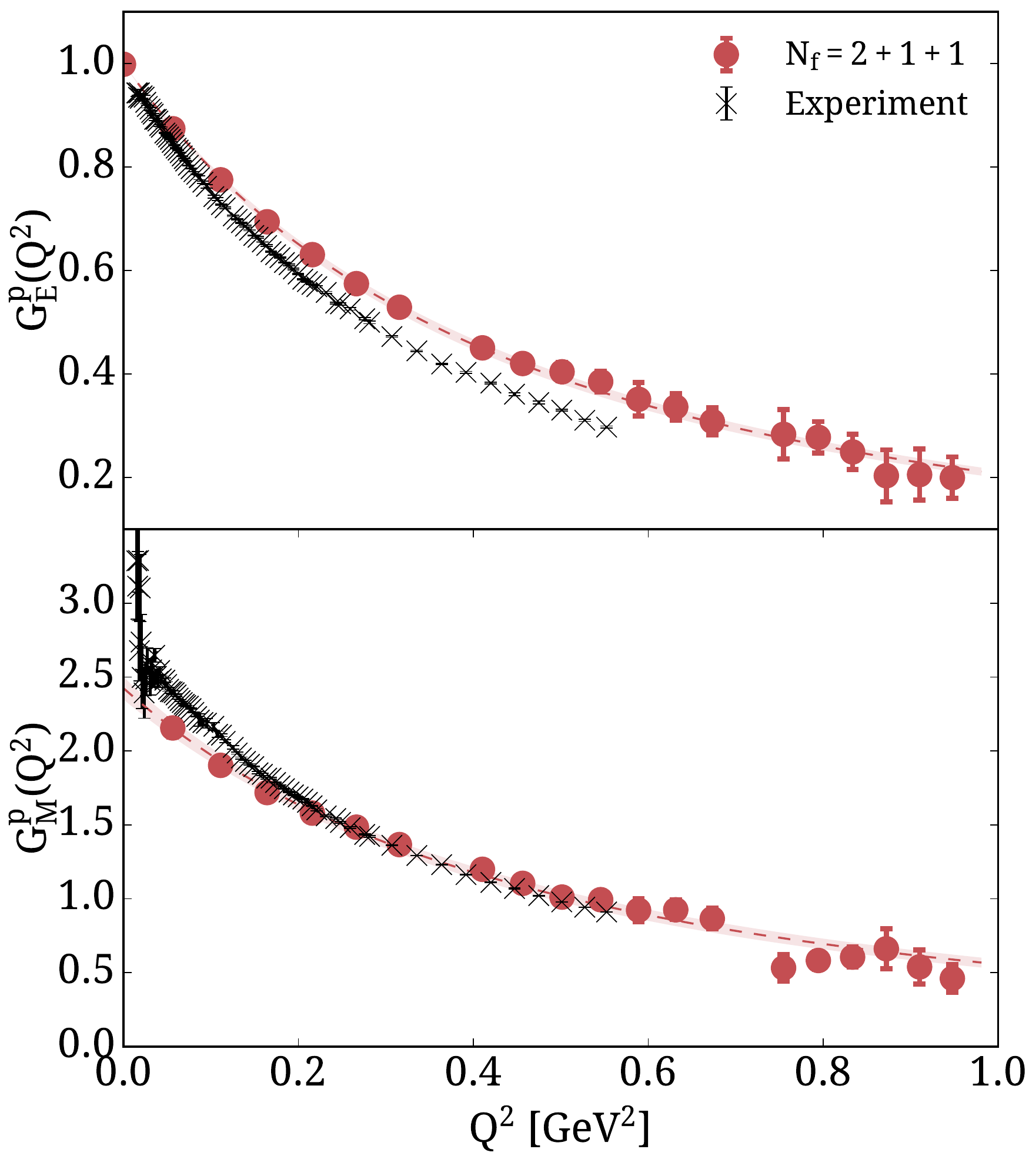}\hspace{4mm}
\includegraphics[width=3.0in]{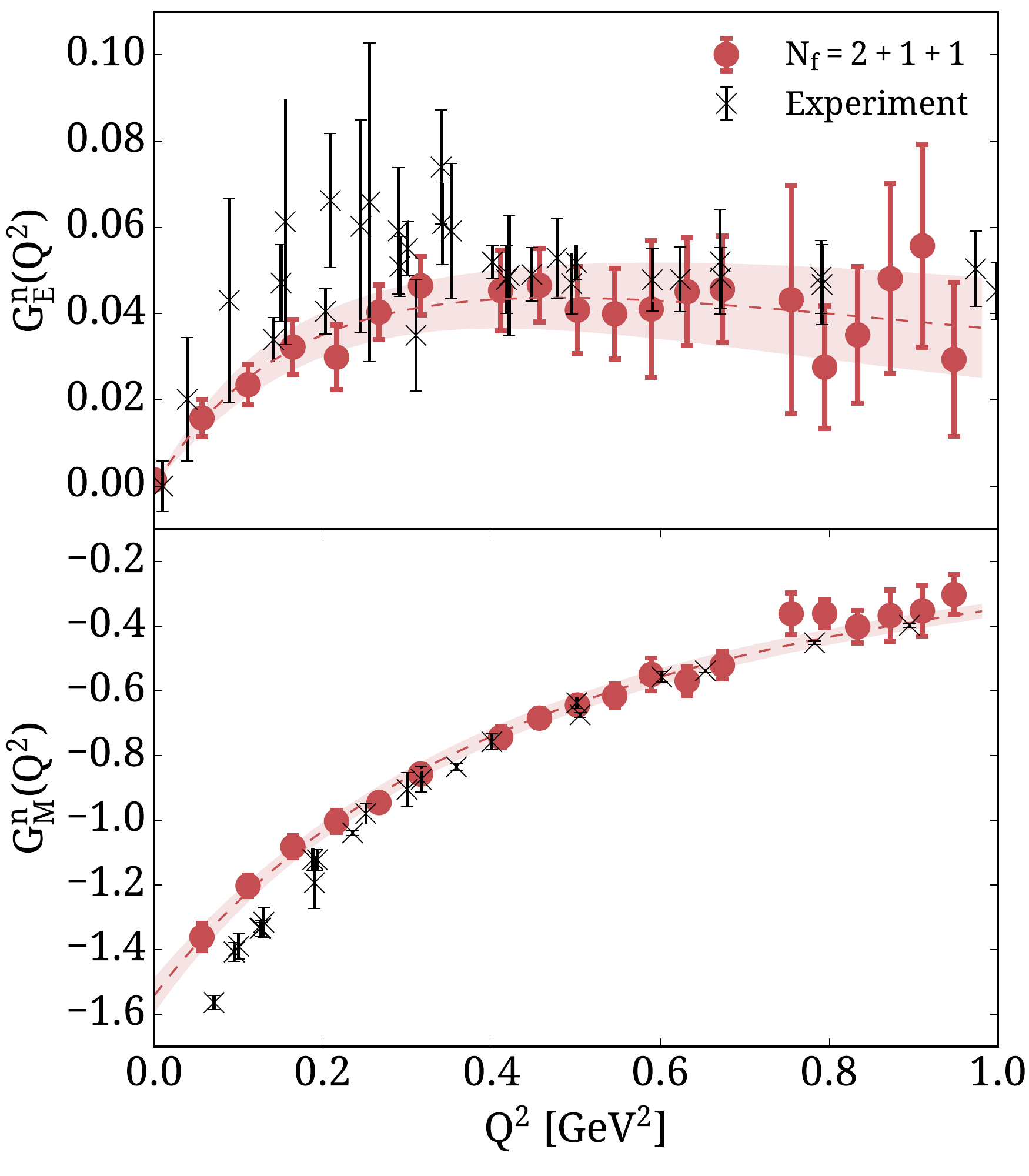}
\caption{Electric (top) and magnetic (bottom) form factors of the proton (left)
and neutron (right) from Ref.~\cite{formfactors}, compared to experiment.}
\label{fig:GEMpn}
\end{figure*}

A recent study of the proton and neutron electromagnetic form factors
in lattice QCD was presented in Ref.~\cite{formfactors}.
One ensemble using an $N_f=2+1+1$ twisted mass action with $m_\pi=130~{\rm MeV}$,
and two ensembles using an $N_f=2$ twisted mass action with $m_{\pi}=130~{\rm MeV}$ 
and two volumes $Lm_\pi\sim 3$ and $Lm_\pi\sim 4$ were utilized.  An
unprecedented precision of the disconnected diagram contributions was achieved
using hierarchical probing, low mode deflation, and large numbers of
smeared point sources to reduce gauge noise.  The results demonstrated that
the disconnected diagrams have nonnegligible effects.  The study included
a thorough investigation of excited-state contamination, but further study of 
finite-volume effects at low $Q^2$ is needed.  Their $N_f=2+1+1$ results 
are compared to experiment in Fig.~\ref{fig:GEMpn}.

%
%
%




\section{Light-cone parton distribution function}
A first determination of the unpolarized helicity parton distribution
  function (PDF) at the physical point with nonperturbative renormalization
  and large momenta treated was presented in Ref.~\cite{parton}.
Extracting PDFs from their moments is impractical, so these authors
 used a clever method proposed by Ji\cite{JiMethod} with subsequent refinements.
 First, they computed spatial correlations between boosted nucleon states,
 then carried out Fourier transforms to produce quantities known as quasi-PDFs,
 then finally took the infinite-momentum limit via a refined matching procedure.
 So-called target mass corrections were employed, as well as a
 renormalization scheme for the Wilson line operators.
 
\begin{figure*}
\centering
\includegraphics[width=3.0in]{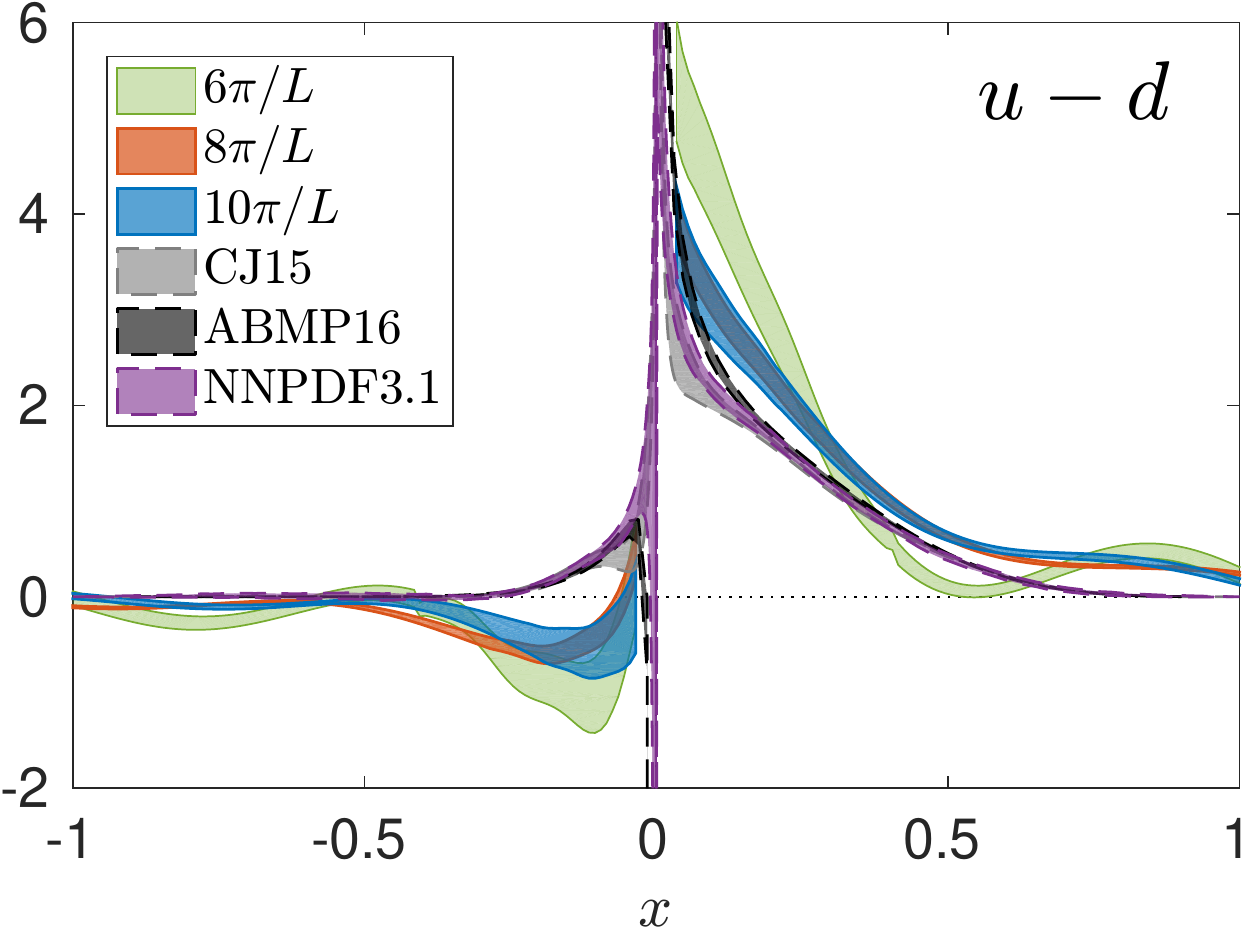}\hspace{4mm}
\includegraphics[width=3.0in]{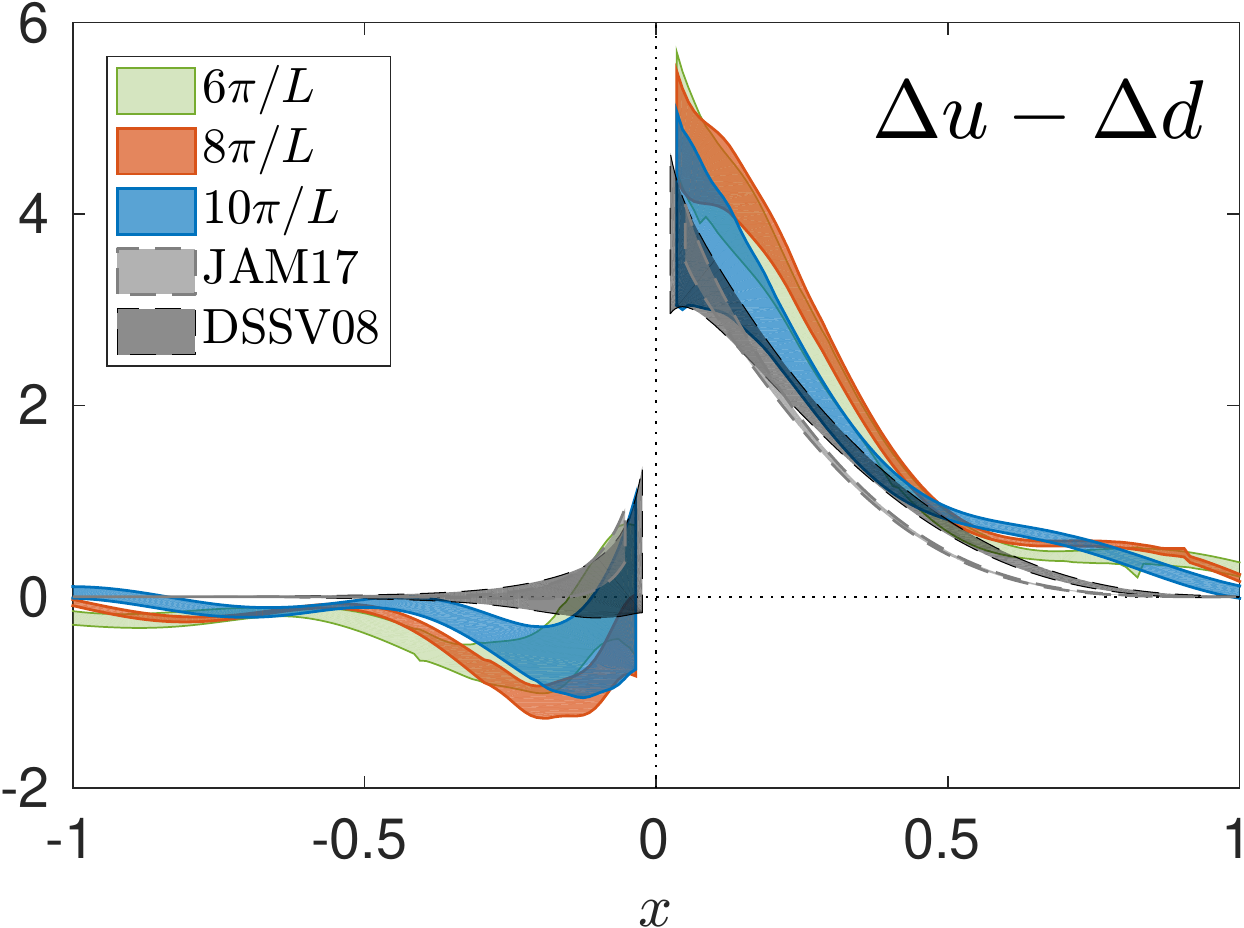}
\caption{Unpolarized (left) and polarized (right) parton distribution functions 
for three momenta compared
to some  phenomenological curves.  Results are from Ref.~\cite{parton}.}
\label{fig:pdfs}
\end{figure*}

 Results were obtained on one ensemble using a $48^3\times 96$ lattice with
 a twisted mass $N_f=2$ action with lattice spacing $a=0.0938(3)(2)~{\rm fm}$ and
  $m_\pi L = 2.98(1)$ at the physical point.
 Both unpolarized and polarized PDFs for three momenta were compared to some 
 phenomenological curves, shown in Fig.~\ref{fig:pdfs}.
 
\section{Excited baryon states}
In finite volume, the stationary-state energies are discrete due to
momentum quantization, so temporal correlation matrices have the spectral
representation   
   \beq
   C_{ij}(t) = \sum_n Z_i^{(n)} Z_j^{(n)\ast}\ e^{-E_n t},
   \qquad\quad Z_j^{(n)}=  \me{0}{O_j}{n},
   \eeq
neglecting wrap-around corrections.  It is not practical to do fits using
the above form, so diagonalization methods with single- and two-exponential
fits are employed to extract some number of low-lying states.

To access excited baryon (and meson) states reliably, it is necessary to
extract the energies of the multi-hadron states that are lower-lying than
the excited baryons of interest.  This requires evaluating correlators
involving multi-hadron operators.  Good multi-hadron operators involve
combining good individual hadron operators which separately have 
well-defined momenta.  With such operators, the usual point-to-all trick,
which exploits translational invariance,
cannot be utilized to drastically reduce the number of quark propagator sources
that are needed, so an overly large number of Dirac matrix inversions are 
required.  One solution to this is the stochastic LapH method\cite{laph}
which estimates the entire matrix inverse, or a large portion of it, using an
additional Monte Carlo calculation, exploiting the Laplacian-Heaviside quark 
smearing and various noise dilution projectors for variance reduction.

\begin{figure*}
\centering
\includegraphics[width=2.8in]{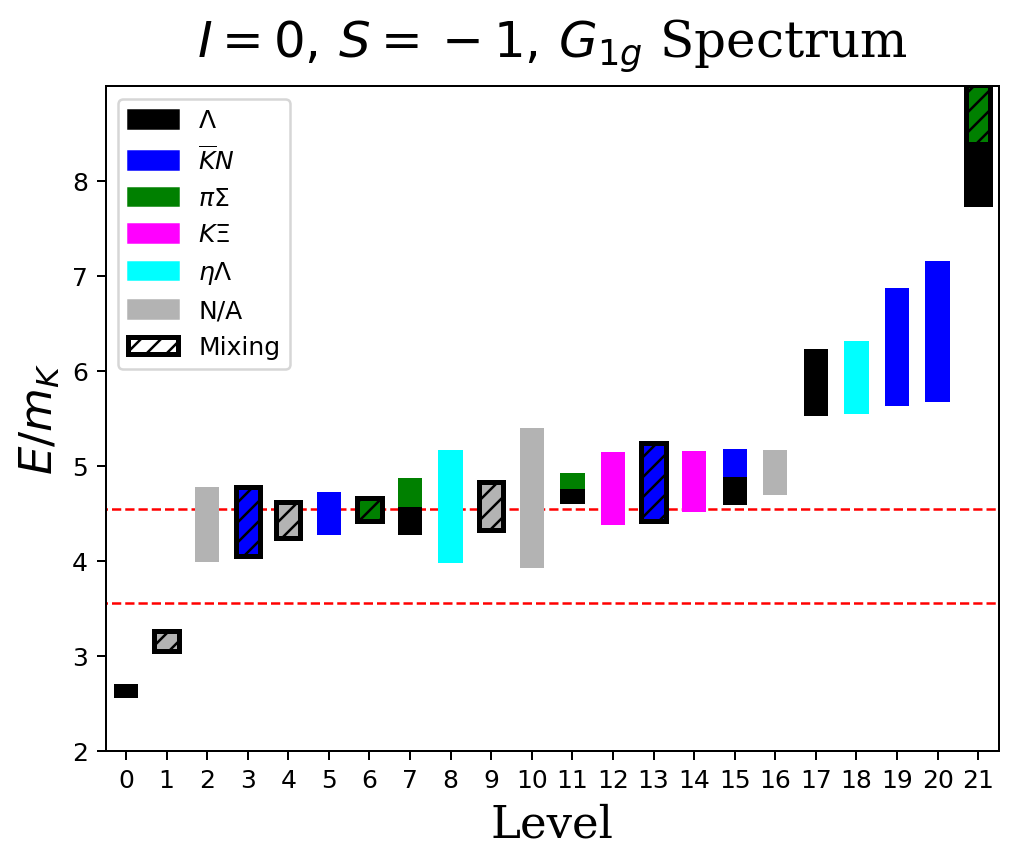}
\includegraphics[width=3.5in]{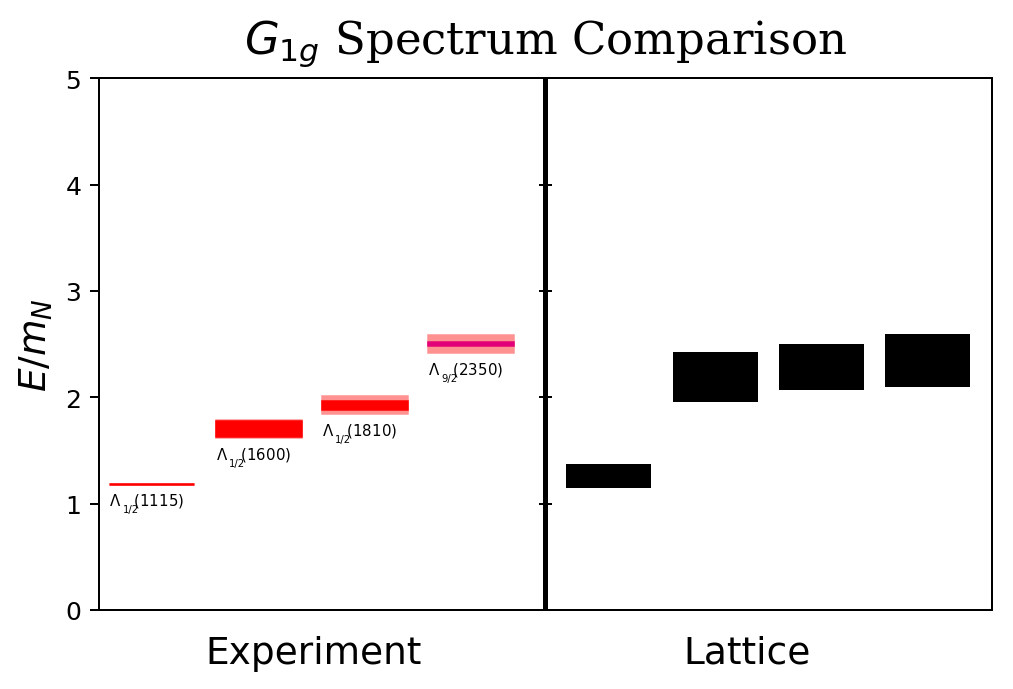}\\
\includegraphics[width=2.8in]{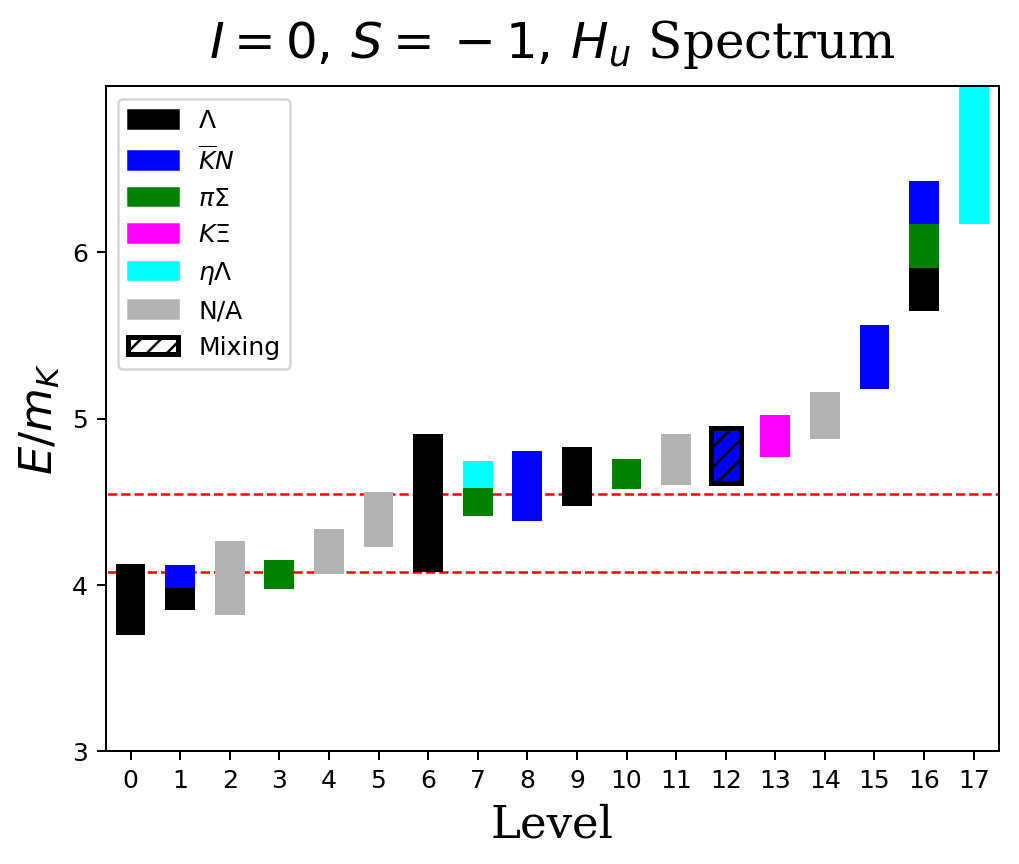}
\includegraphics[width=3.5in]{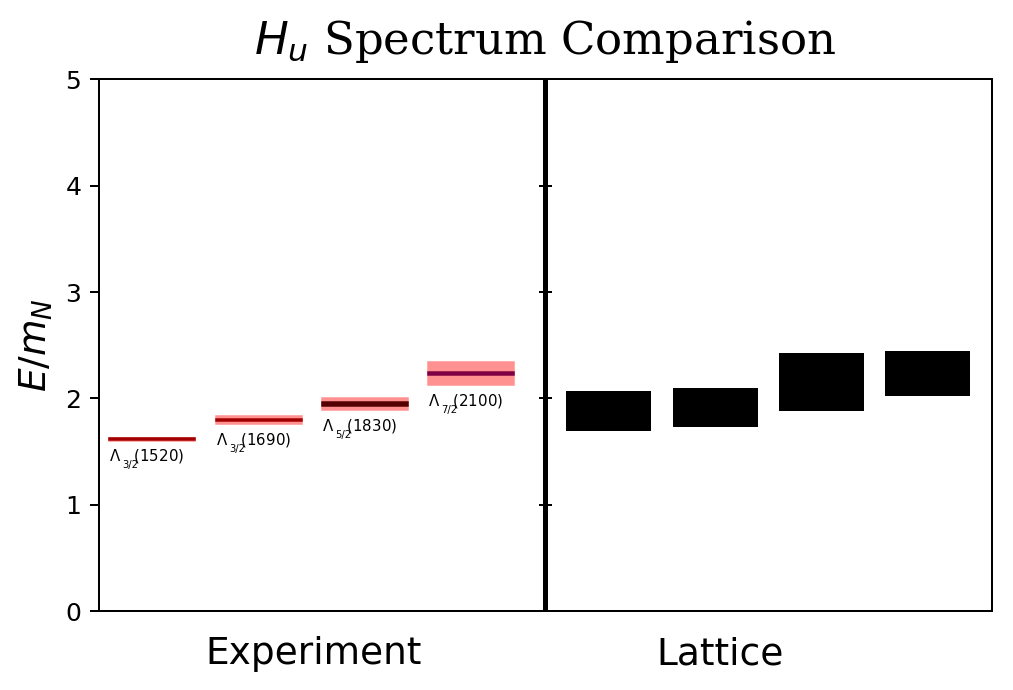}
\caption{(Left) Stationary state energies in the $I=0,\ S=-1$ baryonic
flavor sector for a $32^3\times 256$ anisotropic lattice with 
$m_\pi\sim 240$~MeV.  The $G_{1g}$ channel (even parity containing
the spin-$\frac{1}{2}$ states) and the $H_u$ channel (odd parity 
including the spin-$\frac{3}{2}$ states) are shown as ratios over
the kaon mass $m_K$.  Different colors indicate the different properties
of the states as deduced by the fitted overlap factors $Z_j^{(n)}$.
(Right) The single-baryon dominated levels are compared to experiment
as ratios over the nucleon mass $m_N$.  Results are from Ref.~\cite{fallicathesis}.}
\label{fig:excitedbaryons}
\end{figure*}

Excited $\Lambda$ baryons were studied somewhat recently in 
Ref.~\cite{fallicathesis}.  A small sample of the results obtained is
shown in Fig.~\ref{fig:excitedbaryons}.  

\section{Scattering amplitudes from lattice QCD}
Finite-volume energies $E$ in lattice QCD are related to the 
infinite-volume $S$ matrix\cite{luscher}.  Utilizing such relations to
obtain scattering amplitudes is often termed the L\"uscher method.
Introduce the $K$-matrix as usual,
\beq
   S = (1+iK)(1-iK)^{-1} = (1-iK)^{-1}(1+iK).
\eeq
In the $JLSa$ basis, for total angular momentum $J$, orbital angular momentum
$L$, intrinsic spin $S$, and species channel $a$, 
introduce
\beq
 K^{-1}_{L'S'a';\ LSa}(E)=q_{{\rm cm},a'}^{-L'-\frac{1}{2}}
  \ {\widetilde{K}}^{-1}_{L'S'a';\ LSa}(\Ecm)
  \ q_{{\rm cm},a}^{-L-\frac{1}{2}},
\eeq
then below 3-particle thresholds, there is a quantization condition
\beq
\det(1-B^{(\Pvec)}\widetilde{K})=\det(1-\widetilde{K}B^{(\Pvec)})=0
\eeq
or
\beq
  \det(\widetilde{K}^{-1}-B^{(\Pvec)})=0
\eeq
where the Hermitian ``box matrix'' $B^{(\Pvec)}$ encodes the effects of 
the cubic finite-volume. Details about the box matrix may be found
in Ref.~\cite{boxmatrix}.

The quantization condition relates a single energy $E$ to the entire $K$-matrix,
so one cannot solve for the $K$-matrix, except in a single channel for a single 
partial wave.  Thus, we must approximate the $K$-matrix with functions depending 
on a handful of fit parameters.  We then obtain estimates of the fit parameters 
using many different energies. The quantization condition involves an 
infinite-dimensional determinant.  We make the condition practical by first
transforming to a block-diagonal basis, then truncating in orbital angular momentum.
Meson-meson scattering studies are becoming mature, whereas only a few meson-baryon
scattering investigations have been attempted.  Baryon-baryon scattering studies
are currently gestating.

\begin{figure}
\centering
\includegraphics[width=3.0in]{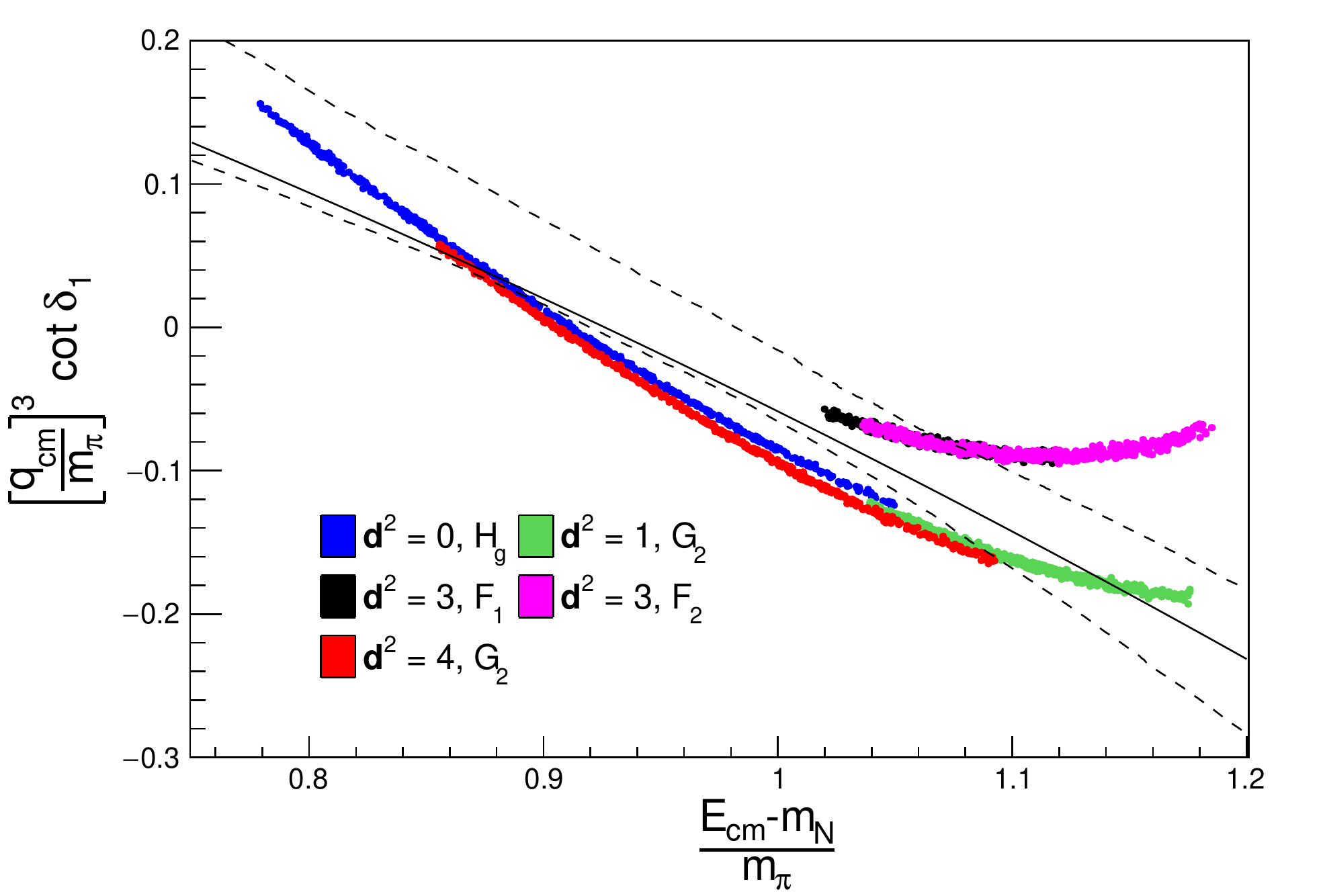}
\caption{$P$-wave $N\pi$ scattering phase shift against center-of-mass energy
showing the $\Delta$ resonance where the $\cot\delta_1$ crosses zero. Results are
from Ref.~\cite{delta}.}
\label{fig:delta}
\end{figure}

The decay $\Delta(1232)\rightarrow N\pi$ was recently studied 
in Ref.~\cite{delta}.  Only the $L=1$ partial wave was included.  Results
were obtained on a large $48^3\times 128$ isotropic lattice with 
$m_\pi \approx 280$~MeV and $a\sim 0.076$~fm.  Their determination of the
scattering phase shift is shown in Fig.~\ref{fig:delta}.  A Breit-Wigner fit gave 
$m_\Delta/m_\pi = 4.738(47)$ and $g_{\Delta N\pi}=19.0(4.7)$. Note that 
experiment yields $g_{\Delta N\pi}\sim 16.9$.  

\begin{figure}
\centering
\includegraphics[width=2.4in]{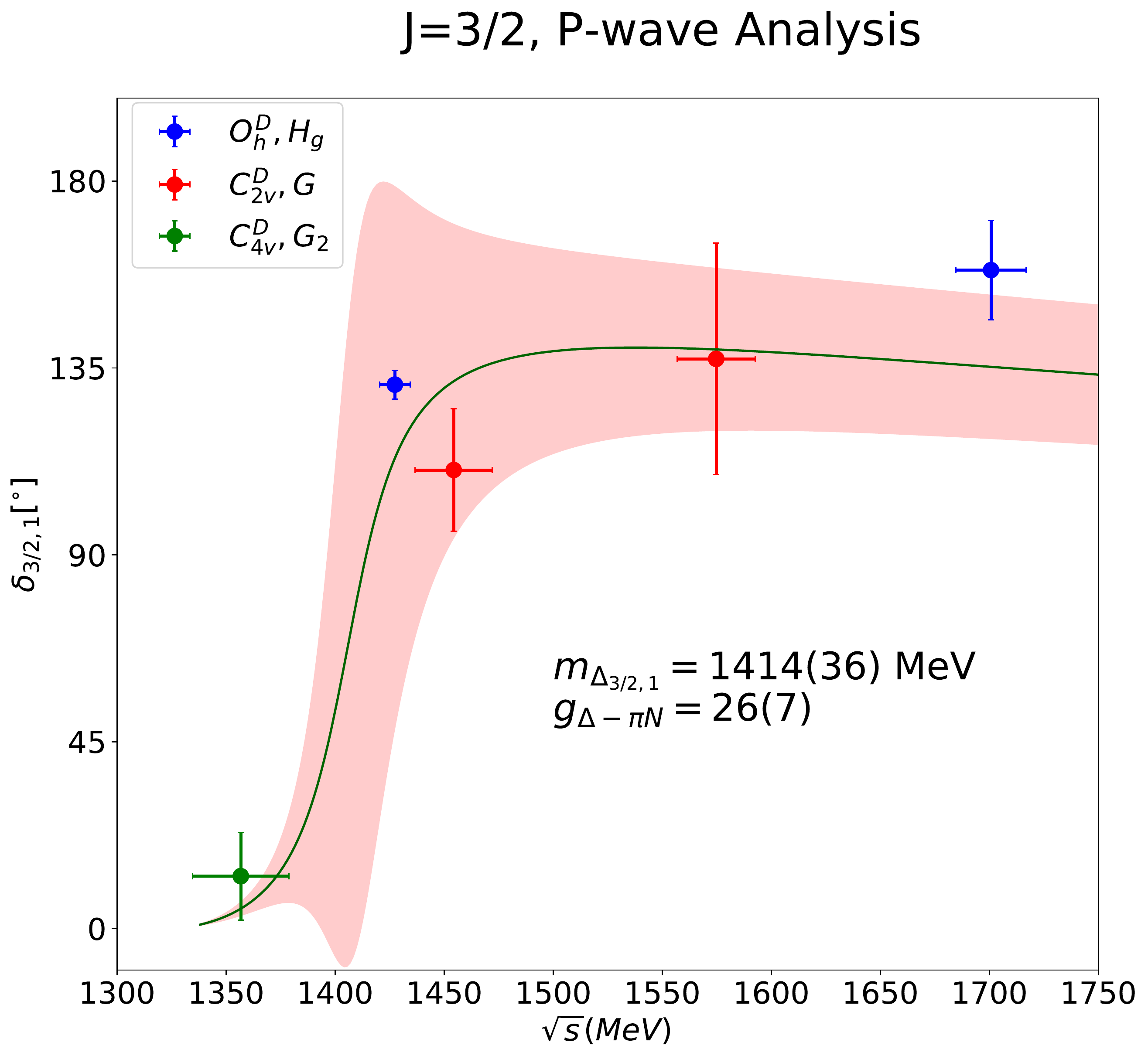}
\caption{$P$-wave $N\pi$ scattering phase shift against the Mandelstam
variable $\sqrt{s}$ showing the $\Delta$ resonance where the phase shift rises
dramatically. Results are
from Ref.~\cite{delta2}.}
\label{fig:delta2}
\end{figure}

Preliminary results for the $\Delta$ resonance in another recent study using a lattice
with length $L=2.8~\rm{fm}$, spacing $ a=0.116~\rm{fm}$ and pion mass
$ m_\pi=260~{\rm MeV}$ has appeared in Ref.~\cite{delta2}.  No slice-to-slice
propagators were used, three total momenta were studied, the ground
and excited states were extracted, and their analysis included only
a single partial wave.  Their determination of the scattering phase shift
is shown in Fig.~\ref{fig:delta2}.

\begin{figure}
\centering
\includegraphics[width=3.4in]{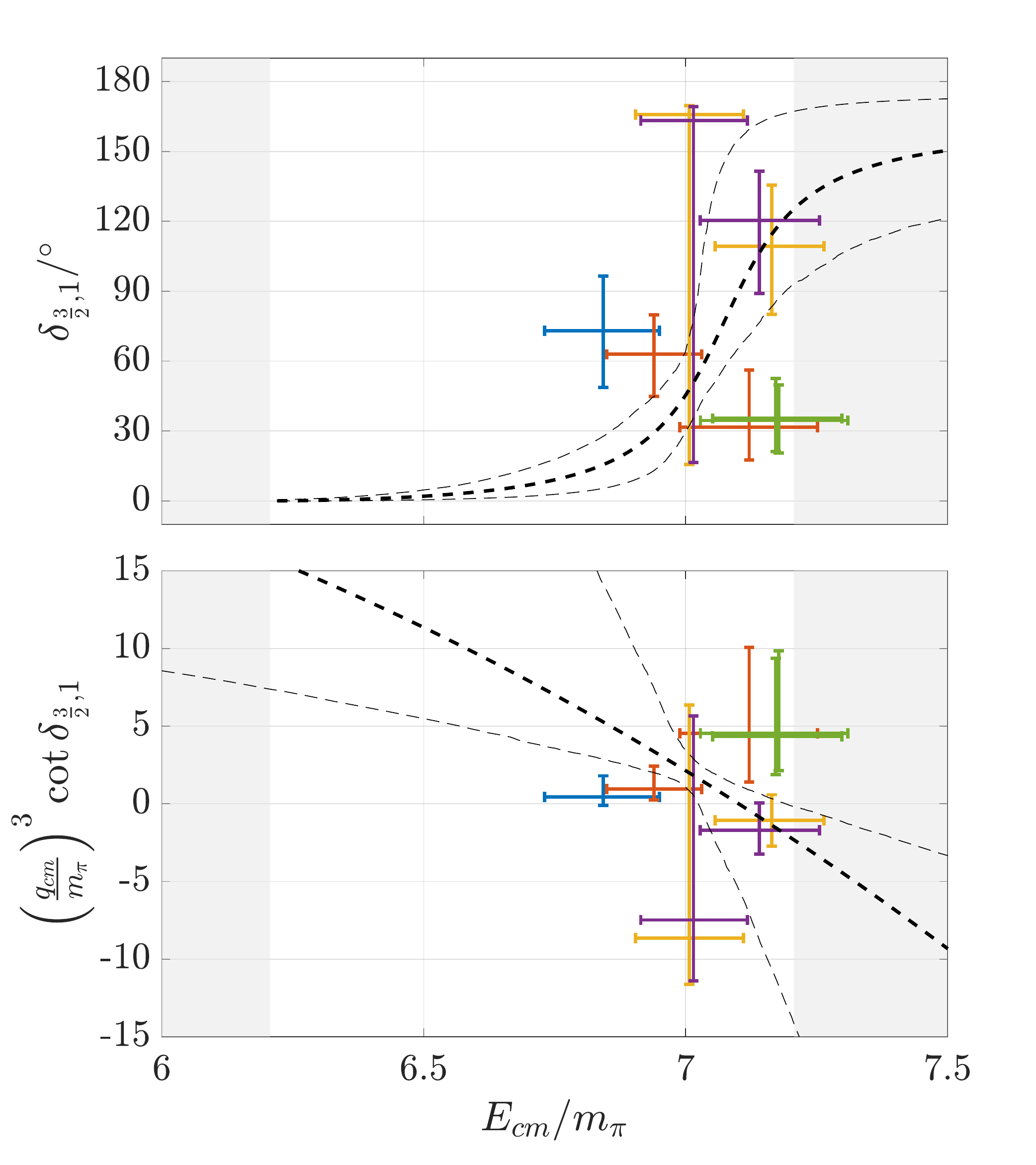}
\caption{$N\pi$ scattering phase shifts near the $\Delta$ resonance
from Ref.~\cite{delta3} against the center-of-mass energy as a ratio
over the pion mass. The top plot shows the phase shift itself, and the
bottom plot shows the cotangent of the phase shift multiplied by
threshold factors.}
\label{fig:delta3}
\end{figure}

\begin{figure}
\centering
\includegraphics[width=3.4in]{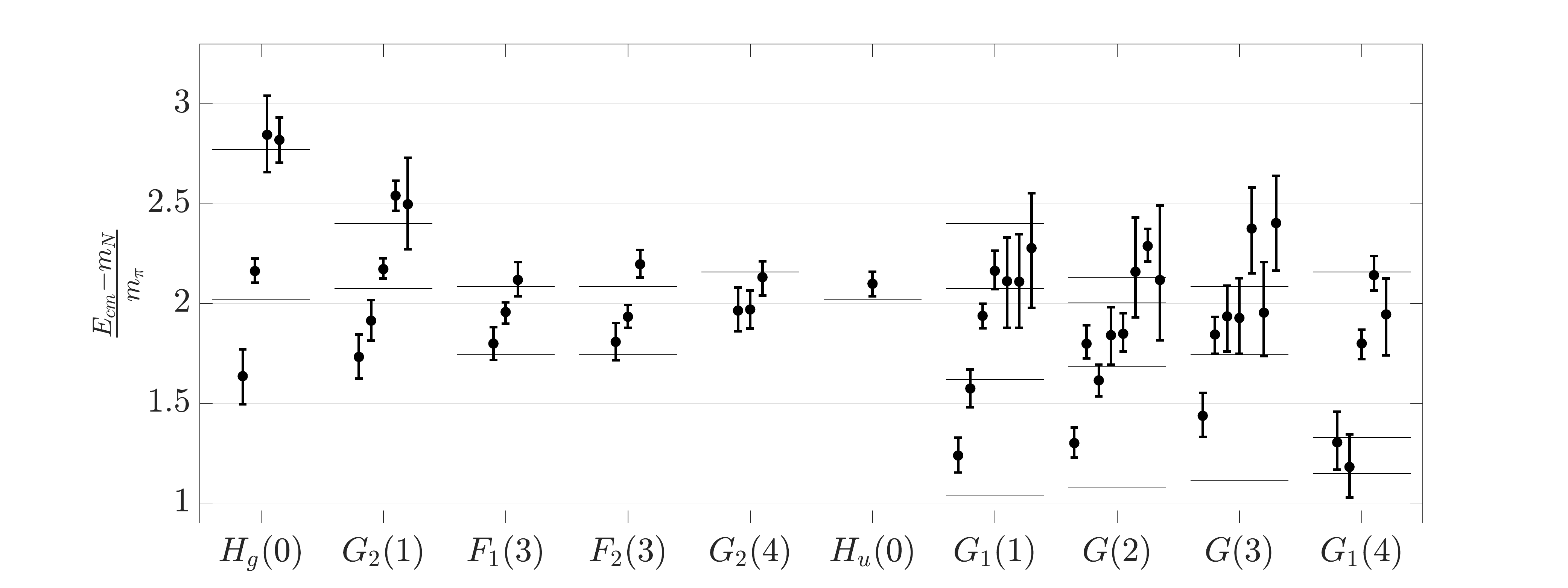}
\caption{The energy spectrum of finite-volume states used in our
most recent study in Ref.~\cite{delta3} of the $\Delta$ resonance.
Differences of the center-of-mass energies from the nucleon mass
as a ratio over the pion mass are shown for the different irreps
and total momenta used. Horizontal lines show the non-interacting
energies.}
\label{fig:deltaspect}
\end{figure}

Preliminary results from our most recent study\cite{delta3} of the 
$\Delta$ resonance are shown in Fig.~\ref{fig:delta3}.  The lattice
length, lattice spacing, and pion mass are 
 $L=4.2~\rm{fm},$ $a=0.065~\rm{fm}$, $m_\pi=200~{\rm MeV}$.
Five total momenta have been used, and both ground and excited states
were extracted.  We expect to reduce statistical errors by a factor of
6 in our final results.  The finite volume spectrum which produced
these results is shown in Fig.~\ref{fig:deltaspect}.  
Fits included irreps which mix the $S$ and $P$ waves and relied
on the automated determination of $B$-matrix elements from Ref.~\cite{boxmatrix}.

\begin{figure}
\centering
\includegraphics[width=3.4in]{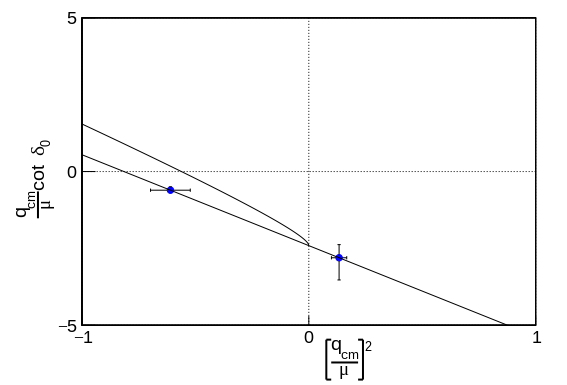}
\caption{Very preliminary results for the $\Sigma \pi$ scattering
phase shift near the $\Lambda(1405)$ resonance against the
center-of-mass momentum as ratio with a reference energy $\mu$.
Result are from Ref.~\cite{lambda}.}
\label{fig:lambda}
\end{figure}

Results for $\Lambda(1405)\rightarrow\Sigma\pi$ will be
presented in the near future in Ref.~\cite{lambda}.  Very preliminary
results on a lattice with 
 $L=3.2~\rm{fm},$ $a=0.065~\rm{fm}$, $m_\pi=280~{\rm MeV}$
are shown in Fig.~\ref{fig:lambda}.

\section{Three last items}

I finish this talk by reporting on three last items of interest.

\begin{figure}
\centering
\includegraphics[width=3.0in]{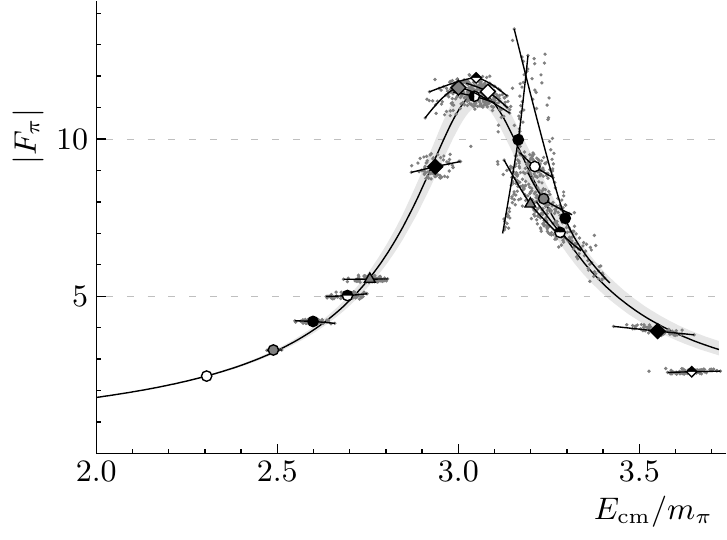}
\caption{
Timelike pion form factor $\vert F_\pi\vert$ again center-of-mass energy over pion
mass $E_{\rm cm}/m_\pi$ using the CLS J303 ensemble on a $64^3\times192$ lattice
with spacing $a=0.050~{\rm fm}$ and 
$m_\pi=260~{\rm MeV}$.  The curve is a fit with a thrice-subtracted dispersion relation.
Results are from Ref.~\cite{pionform}.  This shows the feasibility of future
calculations of baryon form factors.}
\label{fig:pionform}
\end{figure}

A recent determination of the time-like pion form factor appeared
in Ref.~\cite{pionform}.  It was extracted using
 \beqs
 \vert F_\pi(\Ecm)\vert^2 &=& g_\Lambda(\gamma)\left(\qmagcm\frac{\partial\delta_1}{
  \partial \qmagcm}+u\frac{\partial\phi_1^{(\dvec,\Lambda)}}{\partial u}\right)\nonumber\\
 &&\times
\frac{3\pi \Ecm^2}{2\qmagcm^5 L^3}\vert\langle 0\vert V^{(\dvec,\Lambda)}
 \vert \dvec\Lambda n\rangle\vert^2,\nonumber
\eeqs
where
\[ \gamma=\frac{E}{\Ecm},\quad u=\frac{L\qmagcm}{2\pi},\quad
  g_\Lambda(\gamma)=\left\{\begin{array}{ll} \gamma^{-1},&\Lambda=A_1^+\\
  \gamma, & \mbox{otherwise}\end{array}\right.
\]
$\delta_1$ is the physical phase shift, and the
pseudophase $\phi_1^{(\dvec,\Lambda)}$ is obtained from
$B_{11}^{(\dvec,\Lambda)}
 = (\qmagcm/m_\pi)^3\cot\phi_1^{(\dvec,\Lambda)}$.
The matrix element 
\[
  V^{(\dvec,\Lambda)}=\sum_\mu b_\mu^{(\dvec,\Lambda)}V_{R,\mu},\qquad
   \sum_\mu b_\mu^{(\dvec,\Lambda)\ast}b_\mu^{(\dvec,\Lambda)}=1,
\]
with
\beqs   
   V_{R,\mu}&=&Z_V(1+ab_Vm_1+a\overline{b}_V{\rm Tr}M_q)\ V_{I,\mu},\nonumber\\
  V_{I,\mu}&=&V_\mu+ac_V\widetilde{\partial}_\nu T_{\mu\nu},\nonumber\eeqs
and
\[
   V^a_\mu=\textstyle\frac{1}{2}\overline{\psi}\gamma_\mu\tau^a\psi,
\qquad \widetilde{\partial}_\nu T^a_{\mu\nu}
 = \frac{1}{2}i\widetilde{\partial}_\nu\overline{\psi}\sigma_{\mu\nu}\tau^a\psi
\]
was computed and used to determine the form factor.  
Results for CLS J303 ensemble on a $64^3\times192$ lattice with spacing $a=0.050~{\rm fm}$ and 
$m_\pi=260~{\rm MeV}$ are shown in Fig.~\ref{fig:pionform}.  The success of this
calculation paves the way for baryon form factor determinations.  A
similar method is now being used for $\Delta$ transition form factors
needed by the Deep Underground Neutrino Experiment (DUNE).

\begin{figure}
\centering
  \includegraphics[width=1.5in,clip]{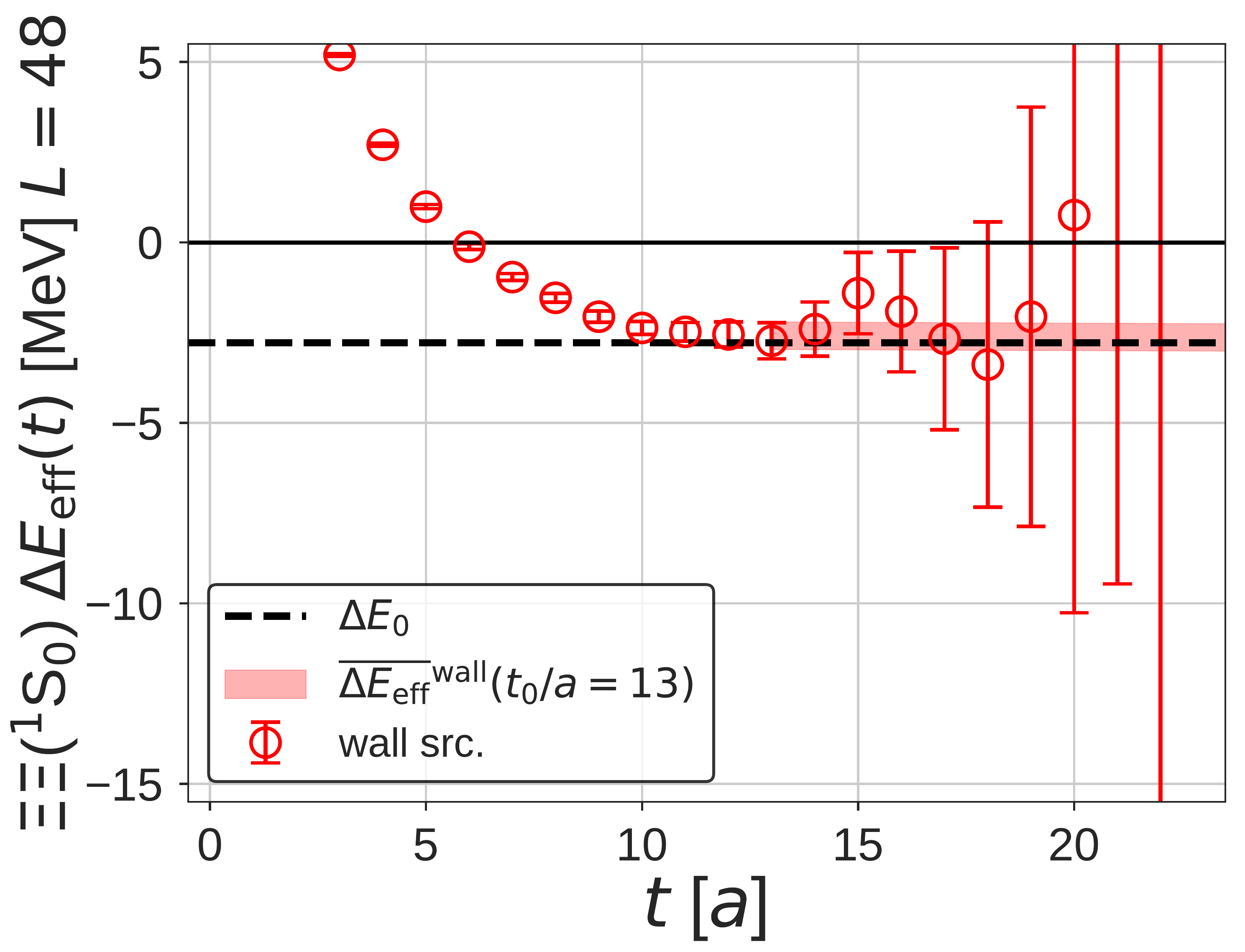}
  \includegraphics[width=1.5in,clip]{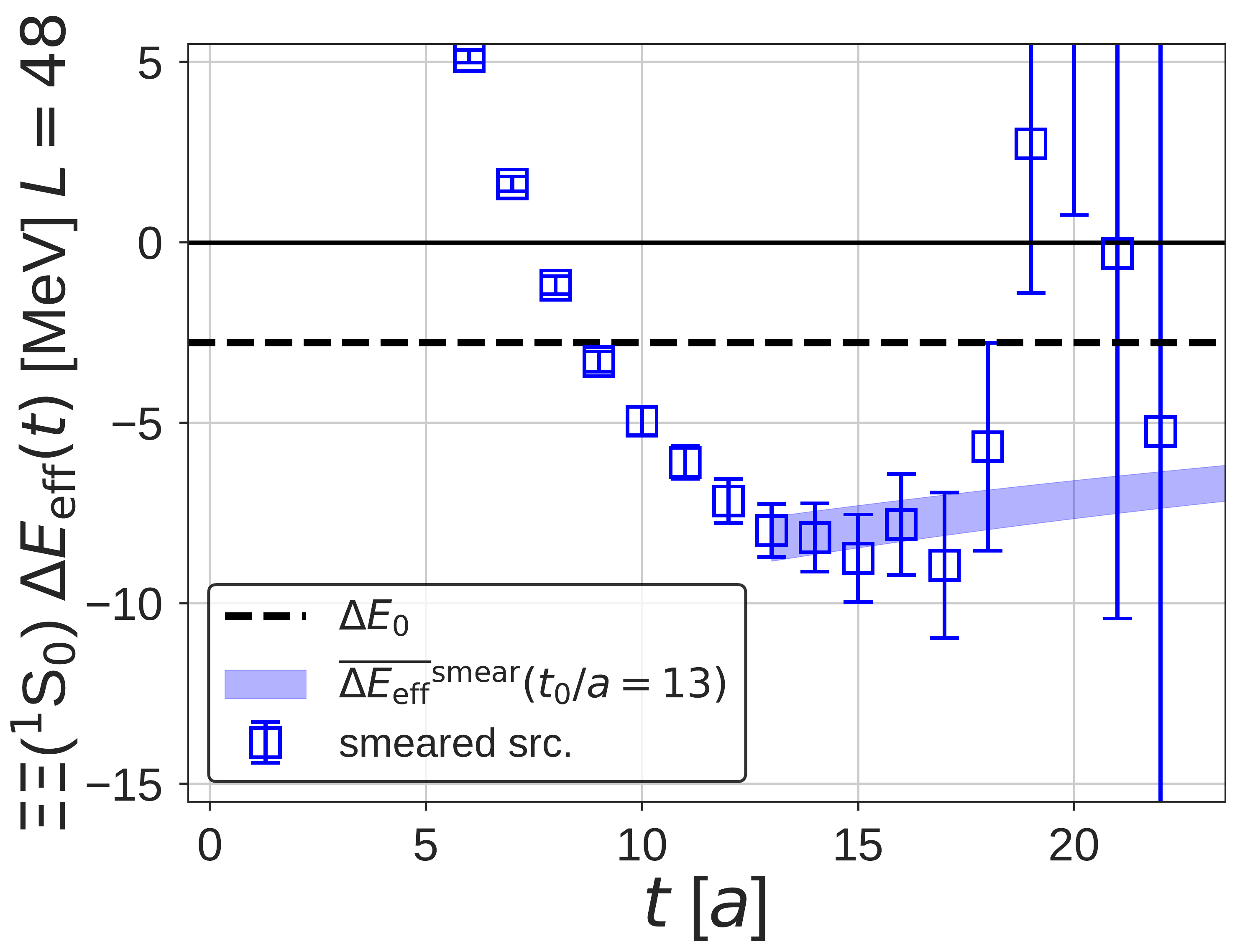}
\caption{Effective masses associated with $\Xi\Xi(^1S_0)$ correlators using
(left) a wall source and (right) a local smeared source.  The dashed line indicates
the asymptotic behavior using the wall source, and this result also agrees with
the HAL QCD method.  The key point, shown in the right plot, is the disagreement 
between the dashed line and the apparent plateau of the effective mass points in 
blue obtained using the local smeared source.  The shaded blue band in the right plot
is the expected large-time behavior. These results are from Ref.~\cite{halqcd}.}
\label{fig:halqcd}
\end{figure}

The  HAL QCD collaboration has extensively studied nucleon-nucleon interactions.
Their method extracts observables from non-local kernels associated
with tempo-spatial correlation functions.  However, a controversy arose when disagreements
of their results with so-called direct methods were found.  A
recent study\cite{halqcd} suggests that this discrepany arises from the
misidentification of energies in the direct method.  The
$\Xi\Xi(^1S_0)$ temporal correlation function was studied in detail, with
pertinent results shown in Fig.~\ref{fig:halqcd}.  In the right plot in this figure,
the key point is the disagreement between the dashed line (the known result) and
the apparent plateau of the effective mass points in blue.  The shaded blue band
is the expected behavior of this function for larger time separations.
Hopefully this resolution
will accelerate progress in baryon-baryon scattering.

\begin{figure}
  \centering
    \includegraphics[width=1.6in]{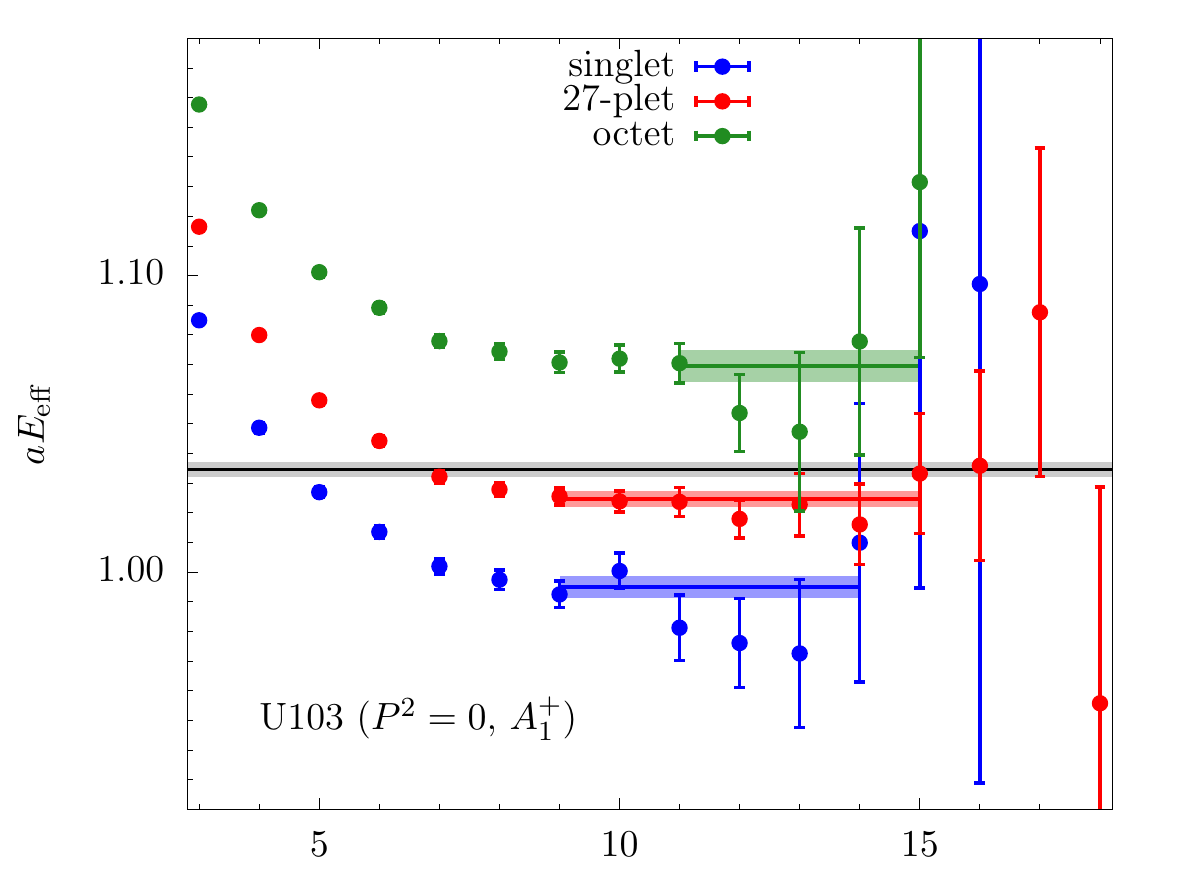}
    \hspace{-0.6cm} 
  \includegraphics[width=1.6in]{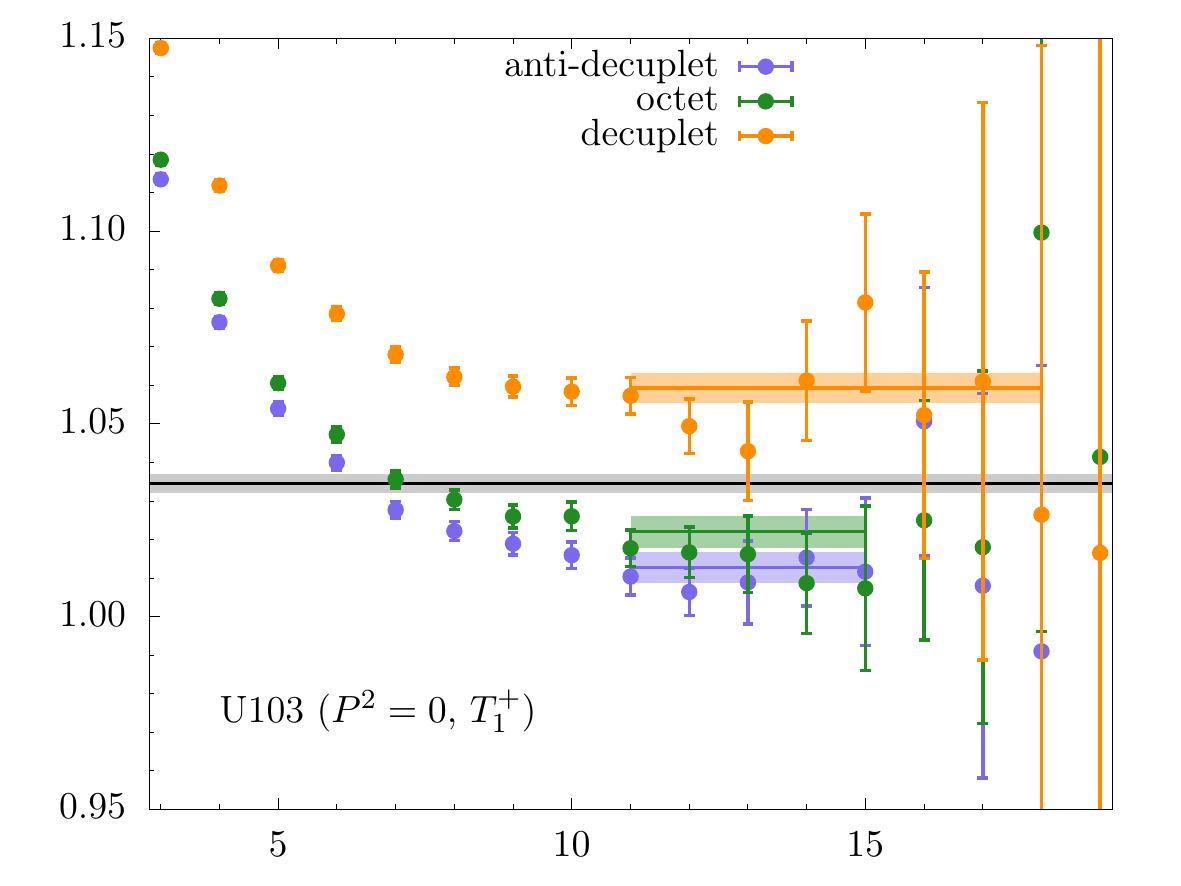}\\[-0.2cm]
    \includegraphics[width=1.6in]{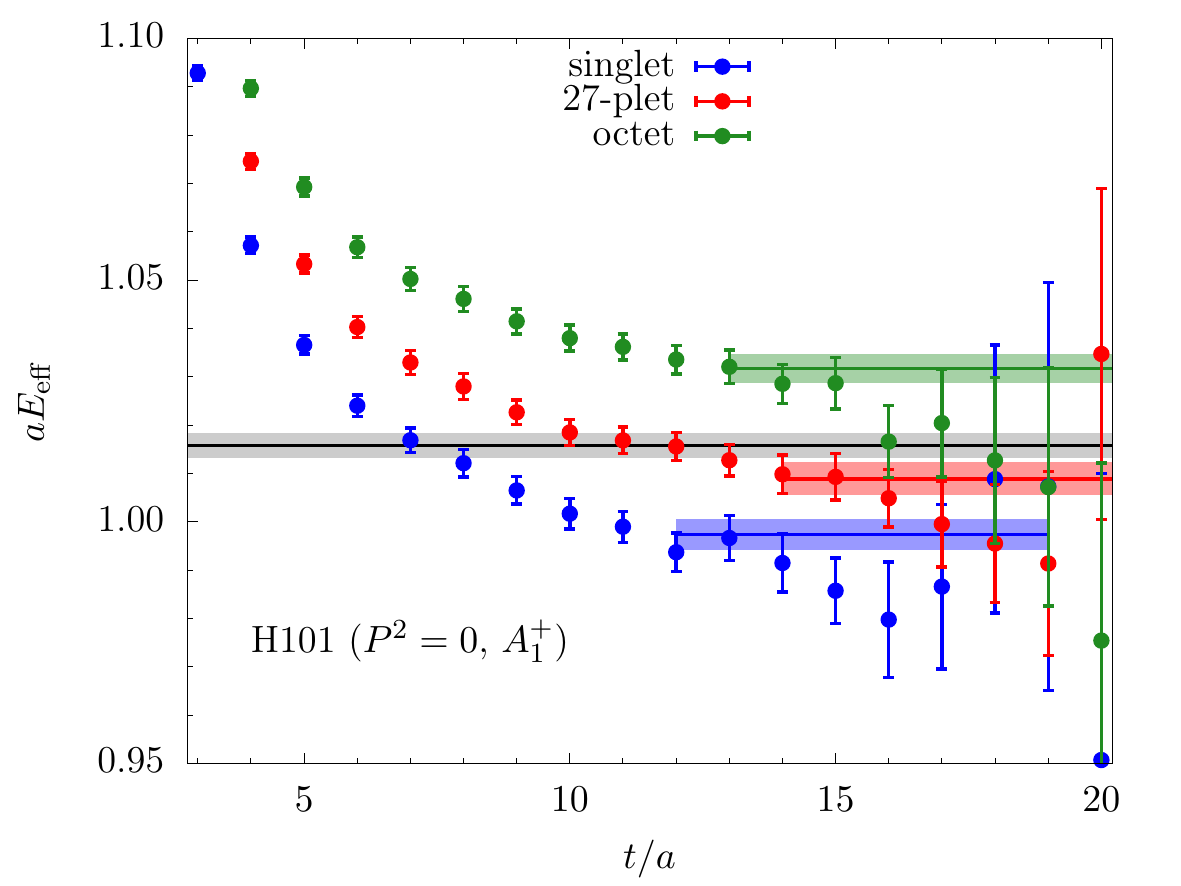}
    \hspace{-0.6cm} 
  \includegraphics[width=1.6in]{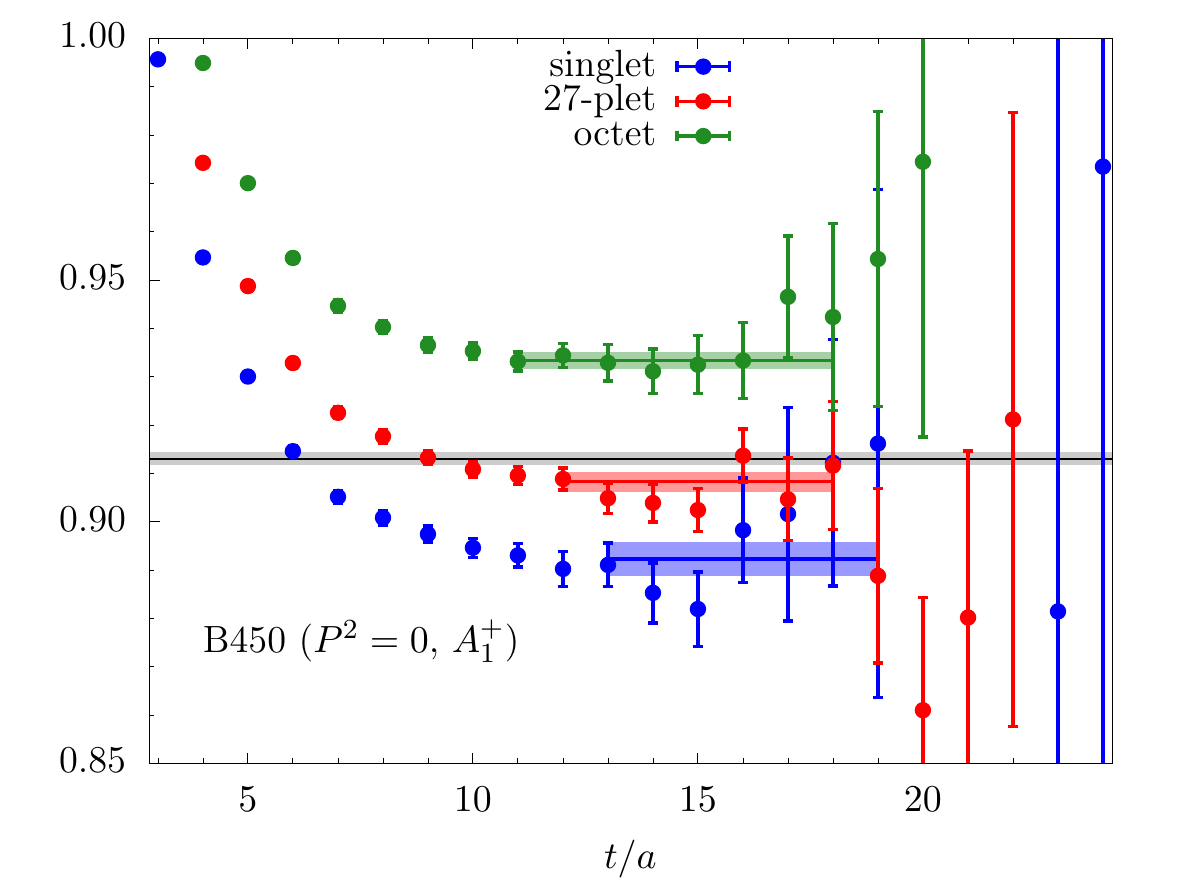}
\caption{Effective masses for spin-0 and spin-1 dibaryon operators of 
 different flavor irreps using 3 ensembles from Ref.~\cite{dibaryon}. Blue points show flavor singlet
 results, the red points show the flavor 27-plet, and the flavor octet is
 shown in green. Results on the
 CLS U103 ensemble in the $A_1^+$ spin-zero irrep are shown in the upper left. 
 The upper right plot shows results on the U103
 ensemble for the $T_1^+$ spin-one irrep. Results in the $A_1^+$ spin-zero irrep
 on the H101 and B450 ensembles are shown in the lower left and lower right 
 plots, respectively.  Horizontal black lines show the two-octet baryon thresholds.
 The $SU(3)$ flavor symmetric point is used, and the lattice volumes are small.}
 \label{fig:dibaryon}
\end{figure}

A recent report on an ongoing study of the $H$-dibaryon was presented
in Ref.~\cite{dibaryon}.  These results were
obtained at the $SU(3)$ flavor symmetric point using well-designed
baryon-baryon operators since a previous study showed that a hexaquark
operator would not saturate the signal.  They found several finite-volume 
energies below the $\Lambda\Lambda$ threshold.  Some of their results
are shown in Fig.~\ref{fig:dibaryon}. A scattering amplitude analysis is
needed to determine if the system is bound or a resonance.  Due to the
small lattices used and the very heavy pion, this must be viewed as
a warm up exercise. Future work on larger lattices and lighter pions 
will involve the stochastic LapH method.

\section{Conclusion}
Highlights from recent computations in lattice QCD involving baryons were presented
in this talk.  How baryons can be studied in lattice QCD was first discussed, followed
by results on the proton mass and spin decompositions, nucleon axial coupling,
the proton and neutron electromagnetic form factors, and light-cone parton 
distribution functions.  Recent works on meson-baryon scattering using the so-called 
L\"uscher method were shown. Key points emphasized were that much better precision
with disconnected diagrams is being achieved, incorporating multi-hadron
operators is now feasible, and more and more studies are being done with
physical quark masses.  

\section{Acknowledgements}
Support from the U.S.~National Science
Foundation under award PHY-1613449 is gratefully acknowledged.

\end{document}